%% file: main.tex
\documentclass[
amsmath,amssymb,
aps,
prd,
floatfix,
twocolumn,
]{revtex4-2}
 
\usepackage{xspace}
\usepackage{graphicx}
\usepackage{hyperref}
\usepackage[T1]{fontenc}
\usepackage{lmodern}
\usepackage{acronym}
\usepackage{xcolor}
\usepackage[caption=false]{subfig}
 
\newcommand{\nn}{\nonumber}

\newcommand{\de}{{\rm d}}

\newcommand{\hu}{\ensuremath{{\rm km \, s^{-1} \, Mpc^{-1}}}\xspace}

\def\eg{{\emph{e.g.~}}}
 
\newcommand{\volume}[1]{\textcolor{red}{\ensuremath{X}}\xspace}

\newcommand{\Hdarklvk}{\ensuremath{67.6^{+13.6}_{-12.7}} \hu \xspace}
\newcommand{\Hlvkcombined}{\ensuremath{71.1^{+9.0}_{-7.1}} \hu \xspace}
\newcommand{\NgwBNS}{{\ensuremath{11044}}\xspace}
\newcommand{\NgwNSBH}{{\ensuremath{9628}}\xspace}
\newcommand{\NgwBBH}{{\ensuremath{46867}}\xspace}
\newcommand{\NgwdetBNS}{{\ensuremath{3}}\xspace}
\newcommand{\NgwdetNSBH}{{\ensuremath{8}}\xspace}
\newcommand{\NgwdetBBH}{{\ensuremath{2324}}\xspace}
\newcommand{\Hbd}{\ensuremath{65.9^{+3.3}_{-2.2}} \hu\xspace}
\newcommand{\Obd}{\ensuremath{0.33^{+0.08}_{-0.06}}\xspace}
\newcommand{\Hbdind}{\ensuremath{64.6^{+3.0}_{-2.7}} \hu\xspace}
\newcommand{\Obdind}{\ensuremath{0.36^{+0.1}_{-0.1}}\xspace}
\newcommand{\Hdnob}{\ensuremath{52^{+15}_{-7}} \hu \xspace}
\newcommand{\Odnob}{\ensuremath{0.57^{+0.3}_{-0.3}}\xspace}
\newcommand{\Hd}{\ensuremath{63^{+10}_{-12}} \hu \xspace}
\newcommand{\Od}{\ensuremath{0.34^{+0.32}_{-0.13}}\xspace}
\newcommand{\Hb}{\ensuremath{64^{+3.5}_{-3.5}} \hu \xspace}
\newcommand{\Ob}{\ensuremath{0.84^{+0.15}_{-0.70}}\xspace}

\begin{document}
%TC:ignore
\input{acronyms}

\preprint{APS/123-QED}
 
\title{Reconsidering the role of bright and dark gravitational-wave standard sirens for cosmology}% Force line breaks with \\
 
\author{Alberto Colombo}
 \email{alberto.colombo@roma1.infn.it}
 \affiliation{INFN, Sezione di Roma, 1-00185 Roma, Italy}%Lines break automatically or can be forced with \\
 \author{Sonia Triscari-Benimav\`o}
  \email{triscaribenimavo.2025706@studenti.uniroma1.it}
 \affiliation{INFN, Sezione di Roma, 1-00185 Roma, Italy}%Lines break automatically or can be forced with \\
 \affiliation{Università di Roma ``La Sapienza'', I-00185 Roma, Italy}%Lines break automatically or can be forced with \\
\author{Simone Mastrogiovanni}%
 \email{simone.mastrogiovanni@roma1.infn.it}
\affiliation{INFN, Sezione di Roma, 1-00185 Roma, Italy
}%

\date{\today}% It is always \today, today,
 
\begin{abstract}
\Ac{GW} standard sirens have emerged as powerful probes of cosmic expansion. They are commonly divided into two classes: bright sirens, observed in coincidence with an \ac{EM} counterpart, and dark sirens, detected without one. Bright sirens have often been considered the main avenue for precision cosmology; however, following the latest observing run of the \ac{LVK} collaboration, the estimates of merger rates of compact objects involving \acp{NS} have decreased.
In this Letter, we reconsider the role that bright and dark sirens will play for precision cosmology. We demonstrate that bright sirens, although detected in small numbers, will play a crucial role in achieving the precision needed to address the $H_0$ tension, primarily by breaking the degeneracy between the Hubble constant ($H_0$) and the matter density parameter ($\Omega_m$) inherent to dark sirens. Alternatively, we show that dark sirens alone can resolve the $H_0$ tension when supplemented with an external prior on $\Omega_m$.
Furthermore, we present a self-consistent statistical framework for \ac{GW} cosmology with bright and dark sirens which have been traditionally treated as independent populations, although they both originate from the same underlying population of \acp{BH} and \acp{NS}. Finally, we find that treating bright and dark sirens as independent populations will not significantly bias GW cosmology in the next LVK observing runs, but can lead to a biased reconstruction of the low-end of the mass spectrum.
\end{abstract}
 
\maketitle

\input{acronyms}
%\tableofcontents
 
%TC:endignore
\section{Introduction}
%\textbf{Letters (length limit: 3750 words)}
 
\Acp{CBC} are self-calibrated \ac{GW} sources \cite{Holz:2005df} for measuring cosmic expansion, but \ac{GW} observations do not provide a direct measure of the source redshift.
Standard sirens are commonly divided into two categories: bright and dark. The former are observed with an \ac{EM} counterpart that allows for a firm identification of their host galaxy and therefore redshift. The latter are not accompanied by the detection of any \ac{EM} counterpart and their localization typically covers a volume of thousands of Mpc$^3$. For dark sirens, it is still possible to statistically infer their redshift with the source-frame mass spectrum \cite{Taylor:2012db, You:2020wju,  Finke:2021aom, Mastrogiovanni:2021wsd, Ye:2021klk, Mancarella:2021ecn, Ezquiaga:2022zkx,MaganaHernandez:2025cnu}, with galaxy catalogues \cite{Schutz:1986gp, Mastrogiovanni:2023emh, Gray:2023wgj}, or cross-correlation with large-scale structure tracers \cite{Oguri:2016dgk, Mukherjee:2020hyn, Mukherjee:2022afz, Bera:2020, Dalang:2024gfk}.
 
The latest \ac{GWTC-5.0} %vecchio\cite{LIGOScientific:2025yae,LIGOScientific:2025hdt, LIGOScientific:2025jau,LIGOScientific:2025pvj,LIGOScientific:2025slb}
\cite{theligoscientificcollaboration2026gwtc50constraintscosmicexpansion,theligoscientificcollaboration2026gwtc50introductionversion50,theligoscientificcollaboration2026gwtc50methodsidentifyingcharacterizing,theligoscientificcollaboration2026gwtc50populationpropertiesmerging}, contains 236 \acp{CBC}, of which only GW170817, a \ac{BNS} merger accompanied by a \ac{sGRB} and a \ac{KN}, is a bright siren \cite{LIGOScientific:2017zic}. GW170817 yielded a measurement of the Hubble constant of $H_0=70^{+12}_{-8} \hu$ \cite{LIGOScientific:2018gmd,Nicolaou:2019cip, Howlett:2019mdh, Mukherjee:2019qmm,Hotokezaka:2018dfi, Palmese:2023beh}, while dark sirens give $H_0=$ \Hdarklvk after marginalizing over the compact object mass spectrum \cite{theligoscientificcollaboration2026gwtc50constraintscosmicexpansion, MaganaHernandez:2025cnu, Pierra:2026ffj, Bertheas:2026odj}. Combining the two measurements while treating the bright and dark siren populations as independent gives \Hlvkcombined. Bright sirens have been proposed as the primary avenue for precision cosmology \cite{Chen:2017rfc}, however, given the lowered merger rate estimates of \acp{CBC} involving a \ac{NS} and the limited detection ranges, it is crucial to reconsider their role in \ac{GW} cosmology.
 
Bright and dark sirens are not independent astrophysical populations: \textit{(i)} the \acp{NS} and \acp{BH} that compose them arise from a common set of formation channels, which jointly determine their mass, spin, and redshift distributions; \textit{(ii)} the nature of the compact objects may be uncertain from the \ac{GW} data; and \textit{(iii)} intrinsically bright sources may be observed as dark because of \ac{EM} selection effects.
Previous studies introduced likelihoods accounting for \ac{EM} selection effects in bright siren cosmology through either semi-parametric detectability models \cite{Chen:2023dgw} or explicit \ac{sGRB} observation models \cite{Mancarella:2024qle}.
 
In this Letter, we instead investigate the consequences of treating bright and dark sirens as independent populations and their complementary contributions to cosmological inference and the reconstruction of the compact object mass spectrum.
Treating bright and dark sirens as independent populations might introduce a systematic bias in the cosmological inference. Furthermore, there are already uncertainties on the nature of some dark sirens with one of the masses falling confidently in the mass gap between \ac{NS} and \ac{BH} \citep{LIGOScientific:2020zkf,LIGOScientific:2024elc}.

We introduce a unified framework for bright and dark siren cosmology that incorporates both outcomes within a single \ac{CBC} population and consistently includes complementary \ac{GW} and \ac{EM} information. We show that the host redshifts of the few bright sirens break the
$H_0-\Omega_m$
degeneracy affecting dark sirens, while dark sirens can reach comparable precision when supplemented by an external prior on $\Omega_m$.
Finally, treating the two populations as independent biases the reconstructed low-mass end of the compact object mass spectrum, although its impact on $H_0$ is modest in our mock realization.
 
\section{Modelling the all-siren likelihood}
 
Let us define the intrinsic binary parameters (source masses and spins) as $\mathbf{i}=\{m_1,m_2,\chi_1,\chi_2\}$ and the extrinsic ones (inclination angle and redshift) as $\mathbf{e}=\{\cos \iota, z\}$. Given \ac{GW} data $x$, host-galaxy redshift data $y$, and population parameters $\Lambda$, including the cosmological parameters, the likelihood for $N_{\rm b}$ bright and $N_{\rm d}$ dark sirens is
 
\begin{widetext}
\begin{align}
\mathcal{L}(\{x\},\{y\}|\Lambda,N_{\rm exp})
& \propto e^{-N_{\rm exp}}\prod_{i}^{N_{\rm d}}
\int \de \mathbf{i}^i \de \mathbf{e}^i \,
\mathcal{L}(x^i|\mathbf{i}^i,\mathbf{e}^i,\Lambda) \nonumber
 \frac{\de N}{\de V_c \de t_{\rm s} \de \mathbf{i} \de \mathbf{e}}
\frac{\de V_c}{\de z} \frac{1}{1+z}
p_{\rm miss}(\mathbf{i}^i,\mathbf{e}^i,\Lambda) \nonumber \\
&\quad \times \prod_{j}^{N_{\rm b}}
\int \de \mathbf{i}^j \de \mathbf{e}^j \,
\mathcal{L}(x^j|\mathbf{i}^j,\mathbf{e}^j,\Lambda)
\mathcal{L}(y^j|z^j) \frac{\de N}{\de V_c \de t_{\rm s} \de \mathbf{i} \de \mathbf{e}}
\frac{\de V_c}{\de z} \frac{1}{1+z}
p_{\rm obs}(\mathbf{i}^j,\mathbf{e}^j,\Lambda).
\label{eq:mainlikelihood}
\end{align}
\end{widetext}
 
 Both terms in the likelihood are functions of the \ac{CBC} merger rate and the \ac{GW} likelihood $\mathcal{L}(x^i|\mathbf{i}^i,\mathbf{e}^i,\Lambda)$, which accounts for measurement uncertainties. The bright siren term includes the redshift information from the host galaxy $\mathcal{L}(y|z)$ and the probability  $p_{\rm obs}(\mathbf{i}^j,\mathbf{e}^j,\Lambda)$ for the \ac{EM} counterpart. The dark siren term is instead weighted by  $p_{\rm miss}(\mathbf{i}^i,\mathbf{e}^i,\Lambda)$, which includes the probability that no counterpart is emitted or that an emitted counterpart is not detected.
 The bright or dark classification therefore emerges from the source parameters and the \ac{EM} selection function, rather than defining two independent populations.
 
 In the limit that all \ac{GW} detections are bright sirens, the likelihood collapses to the one found from first principles in \cite{Mancarella:2024qle}. The likelihood is also consistent (modulo a normalization factor to convert physical probabilities to rates) with the approach proposed in \cite{Chen:2023dgw} to account for selection biases introduced by the EM detection that could bias the $H_0$ inference \cite{Chen:2020dyt}. Conversely, when all events are dark sirens known with certainty to be unable to produce \ac{EM} emission, as for \acp{BBH}, it reduces to the standard dark siren likelihood. We note that, differently from the main result of \cite{Essick:2026nzq}, our likelihood assumes that the probability of \ac{EM} detection (and follow-up by EM observatories), depends on the intrinsic parameters of the source. The complete likelihood derivation is provided in the Supplemental Material.
 
To evaluate $p_{\rm obs}$ and $p_{\rm miss}$, we adopt the semi-analytical framework of Ref.~\cite{Colombo:2025sdm}, considering  \ac{KN}, \ac{GRB} prompt emission or both as possible counterparts. For each binary, we first calculate the dynamical ejecta and accretion disk masses and the ejecta velocity using fits to numerical-relativity simulations. These quantities determine the production and properties of the \ac{KN} and structured-jet \ac{GRB} counterparts. We also impose physically motivated conditions for jet launching \cite{Salafia:2022xjd,Colombo:2022zzp,Colombo:2025sdm}, including the formation of a \ac{HMNS} as a remnant. We classify a \ac{KN} as detectable when its peak $g$-band magnitude satisfies $m_g<26$, corresponding to a ${\sim}180\,{\rm s}$ exposure with the Vera C. Rubin Observatory \cite{LSST:2008ijt,Andreoni:2021epw}. For the \ac{GRB}, we adopt a fiducial bolometric fluence threshold $F_{\rm lim}^{\rm GBM}=3\times10^{-7}\,{\rm erg\,cm^{-2}}$ based on \textit{Fermi}/GBM and an alternative threshold ten times lower to explore improved sensitivity. The emission and observation models are detailed in the End Matter.

\begin{figure}
\centering
\includegraphics[width=1.\linewidth]{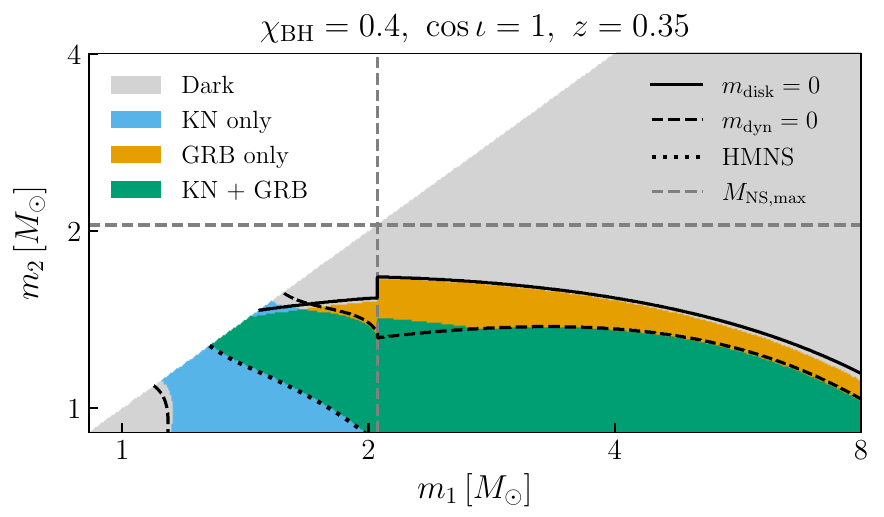}
\caption{EM detectability in the $(m_1, m_2)$ plane, assuming  $\chi_{\rm BH}=0.4$, $\cos\iota=1$, $z=0.35$, the SFHo \ac{EoS} and the improved \textit{Fermi}/GBM sensitivity. Grey corresponds to no detectable emission (dark), while blue, orange, and green indicate detectable \ac{KN}, \ac{GRB} prompt, and joint counterparts, respectively. The grey dashed lines mark the maximum \ac{NS} mass $M_{\rm NS,max}=2.06\,M_\odot$. The black solid, dashed, and dotted curves denote $m_{\rm disk}=0$, $m_{\rm dyn}=0$, and $m_{\rm rem}=1.2\,M_{\rm NS,max}$, respectively, the latter corresponding to the threshold for the formation of a \ac{HMNS}. The corresponding dependence on inclination and redshift is shown in the End Matter.} \label{fig:pdets}
\end{figure}

Figure~\ref{fig:pdets} illustrates how the \ac{EM} selection depends on the intrinsic population. In our deterministic implementation, colored regions have $p_{\rm obs}=1$ for the indicated counterpart, whereas grey regions have $p_{\rm miss}=1$. The boundaries reflect the physical requirements for ejecta, disk, and jet formation, together with the adopted detection thresholds. In particular, binaries may be dark either because they cannot produce a counterpart or because their emission falls below the detection threshold.

\section{Bright and Dark sirens in an O5 scenario}\label{sec:O5scenario}
 
We simulate a set of bright and dark sirens detectable with current \ac{GW} detectors in a fifth observing (O5) run scenario, thus we consider the peak performance that GW cosmology could reach with the currently planned observing runs. We assume a population of \ac{CBC}s with an overall local merger rate of $200 \,{\rm Gpc^{-3} yr^{-1}}$ and increasing in redshift as $(1+z)^{2.7}$. The mass spectrum spans from 1 $M_\odot$ to 300 $M_\odot$, and it is described by an instance of the \textsc{fullpop-4.0} model consistent with results from \cite{LIGOScientific:2025jau}. The maximum mass of a non-rotating \ac{NS} is 2.06 $M_\odot$, and it follows from the choice of the \textsc{SFHo} \ac{EoS}, that is consistent with current observational constraints \cite{Miller:2019cac,Raaijmakers:2021uju}. According to this population model, in one year we expect about \NgwBNS, \NgwNSBH, and \NgwBBH \ac{GW}s from \ac{BNS}, \ac{NSBH}, and \ac{BBH}, respectively, crossing Earth.
 
We assume a 5-detector network composed of three LIGO \cite{LIGOScientific:2014pky} (Livingston, Hanford and India), Virgo \cite{Acernese:2015gua} and KAGRA \cite{Aso:2013eba} at their expected O5 sensitivities. Using \textsc{gwfish} \cite{Dupletsa:2022scg}, we calculate an expected luminosity distance horizon in terms of the binary total detector mass, and we use it to create a proxy for the binary network \ac{SNR}. We consider detected binaries the ones that exceed a network \ac{SNR} threshold of 12, and for each of them, we generate estimates of their luminosity distance, detector masses and inclination angles. Then, we use the \ac{EM} detection model described before to assess the detectability of the two \ac{EM} counterparts if they are emitted.

In one year of observation, we expect to detect \ac{GW}s from \NgwdetBNS \acp{BNS}, \NgwdetNSBH \acp{NSBH}, and \NgwdetBBH \acp{BBH}. Of these detections, all \acp{BBH} are dark sirens. Assuming our fiducial thresholds, a \ac{KN} is detected for all \NgwdetBNS \acp{BNS} and for 5 \acp{NSBH}. If we improve the sensitivity of the \ac{GRB} detector by an order of magnitude, for one of the 5 \ac{NSBH} it is possible to detect also the \ac{GRB} (see Fig.~\ref{fig:simulation_MD1} in the End Matter). We note that these forecasts should be taken with caution, as they are a single realization of a population fit that should include additional uncertainties on the population model. While the \ac{BBH} population is constrained by \NgwdetBBH detections, the \ac{BNS} and \ac{NSBH} populations are inferred from only a few events and therefore remain substantially more uncertain. We also note that the bright siren estimates could be optimistic as we have assumed a spin magnitude of $0.4$ for all \ac{NSBH} and no dependence of the \ac{EM} detection probability on the \ac{GW} sky localization. In particular, the sky localization is expected to reduce the bright sirens detection rate by at least a factor of two\footnote{The factor roughly comes from the day-night cycle and the observation in a single hemisphere}, if the detection rate is dominated by the \ac{KN} detection with optical facilities such as the Vera Rubin LSST.
 
\section{Cosmological and population inference}

\begin{figure*}
    \centering
    \includegraphics[width=0.45\linewidth]{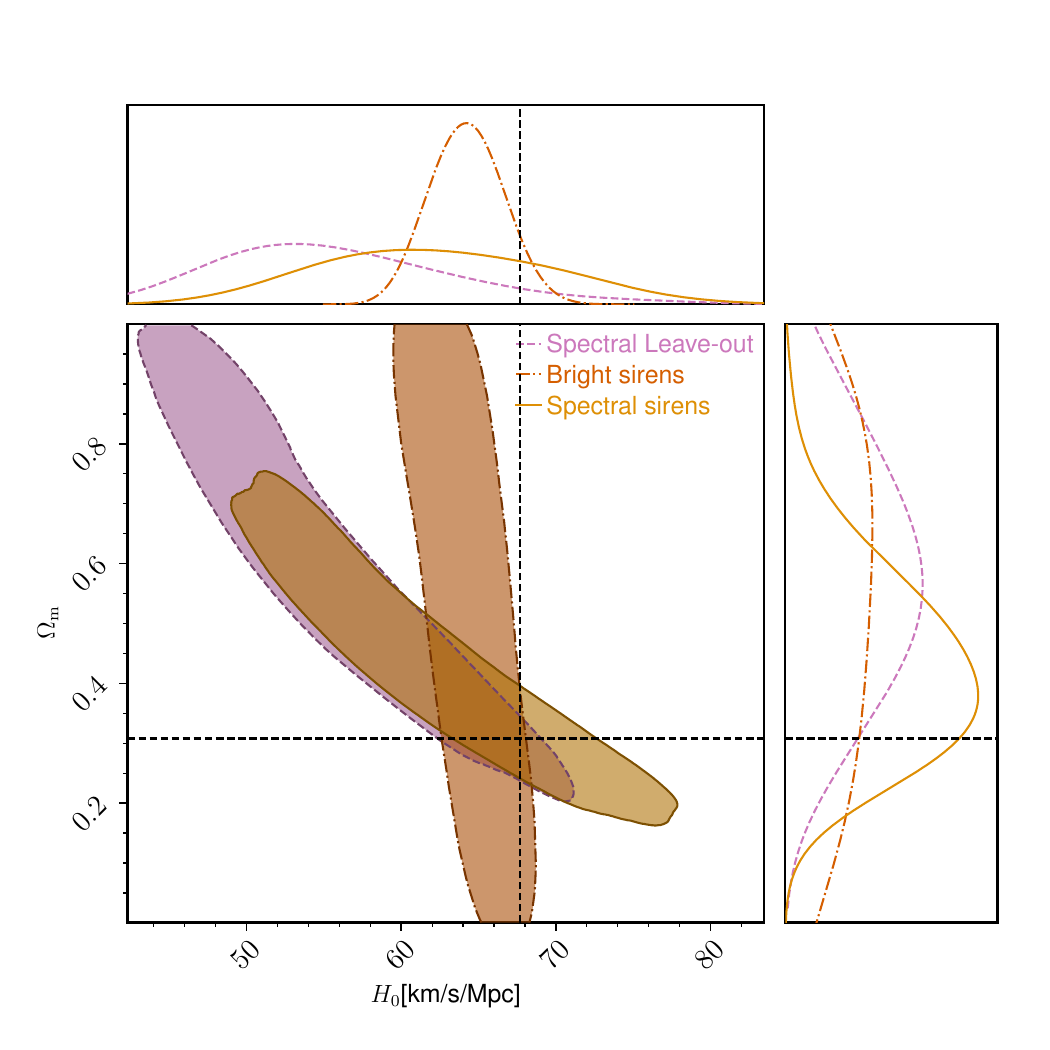}
    \includegraphics[width=0.45\linewidth]{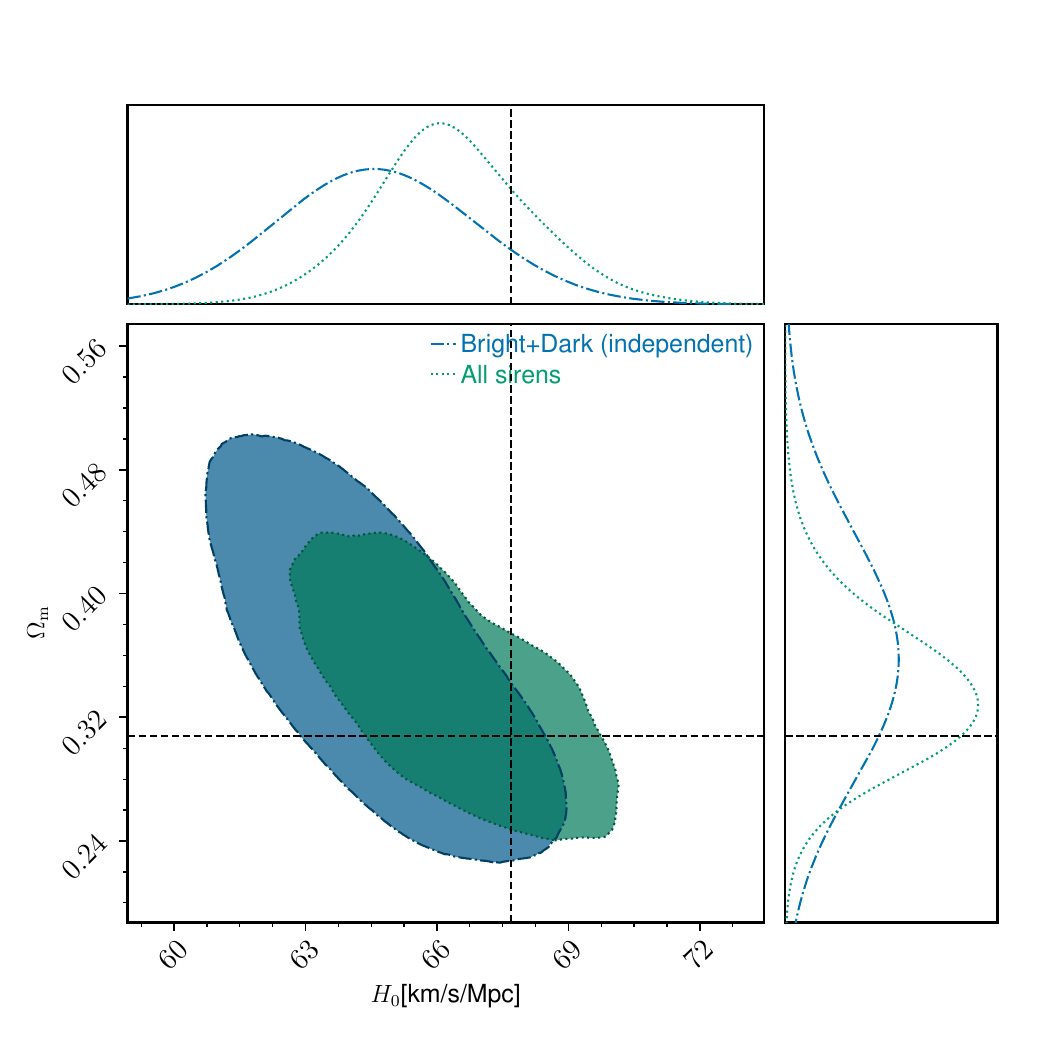}
    \caption{\textit{Left panel:} Posterior distributions of $\Omega_{\rm m}$ and $H_0$ obtained using different inference methods. The contours show the joint 2D posteriors, while the marginalised 1D distributions are reported along the diagonal. The blue, orange, green, and pink contours correspond to the spectral method (all sirens treated as dark), the all-siren method (including both dark and bright sirens), the spectral leave-out method (dark sirens only, excluding bright ones), and the bright sirens method (bright sirens only, excluding dark ones), respectively. The dashed black lines indicate the fiducial cosmological parameters used to generate the mock data. \textit{Right panel:} Same joint posterior but for the full hierarchical Bayesian analysis (green line) and combining bright and dark sirens a posteriori as independent populations (blue).}
    \label{fig:H0_OM0_THESEUS}
\end{figure*}

We use the population of dark and bright sirens described in the previous section for several case studies. In the following, we will use the dataset of bright sirens observable assuming the improved sensitivity of \textit{Fermi}/GBM in order to include a bright siren with an observed \ac{GRB}.
We perform four different analyses sampling over population and cosmological parameters \footnote{see Supplemental Material for more details of the priors}. The first one includes the bright and dark sirens (all sirens method), the second considers all the sirens as dark (spectral sirens method), the third considers only the dark sirens (spectral leave-out), and the fourth considers only the bright sirens (bright sirens method).
The first analysis adopts the likelihood in Eq.~\ref{eq:mainlikelihood}, the second and third the dark siren likelihood as in current literature with no \ac{EM} modeling \cite{theligoscientificcollaboration2026gwtc50constraintscosmicexpansion} and the fourth the bright siren likelihood with no EM modeling \cite{theligoscientificcollaboration2026gwtc50constraintscosmicexpansion}.  For the fourth analysis, we approximate selection biases to scale as $H_0^3$ as we are effectively using only bright sirens that are detectable in the local universe.
In the second analysis we are not intentionally including the redshift information of the host galaxies. The third and fourth analyses are not formally correct, as we are deliberately not including dark/bright sirens without accounting for it in the likelihood; these analyses are useful to assess the impact of dark and bright sirens for cosmological inference, and they are also the current proxies for \ac{GW} cosmology. The third and fourth analyses are then combined, considering the bright and dark sirens populations as independent to mimic the current approach.
 
The left plot of Fig.~\ref{fig:H0_OM0_THESEUS} displays the marginal posteriors on $H_0$ and $\Omega_m$ using populations of dark and bright sirens independently. When we consider all sirens as dark, we obtain $H_0=\Hd$ and $\Omega_{\rm m}=\Od$. The precision \footnote{Note that the fiducial value of $H_0$ is outside the confidence intervals of the marginal distribution due to a projection effect of the degeneracy with $\Omega_{\rm m}$, the joint posterior indeed includes the simulated value.} on these two parameters arises from the degeneracy between $\Omega_{\rm m}$ and $H_0$ that has also been observed for dark sirens in previous studies \cite{2026arXiv260217756T}. Instead, when we consider the dark sirens leaving out the bright sirens, we obtain $H_0=\Hdnob$ and $\Omega_{\rm m}=\Odnob$, and we notice a significant shift in their joint posterior. This shift is due to the reconstructed mass spectrum in the \ac{NS} mass range (see later), and moves the fiducial values of the cosmological parameters at the edge of the $90\%$ credible intervals. Considering only the bright sirens and fixing the population model (we recall that this is an incorrect analysis from the point of view of selection biases) leads to $H_0=\Hb$ and $\Omega_{\rm m}=\Ob$. From Fig.~\ref{fig:H0_OM0_THESEUS}, we notice that bright sirens can be used to break the $H_0 - \Omega_m$ degeneracy for dark sirens.
 
In the right panel of Fig.~\ref{fig:H0_OM0_THESEUS}, we show the joint posteriors on the cosmological parameters when all the sirens are combined. When using the correct statistical framework, we obtain values of \Hbd and \Obd (at 90\%  credible intervals), and therefore we are jointly able to measure $H_0$ and $\Omega_m$ without any systematics. Instead, when combining the dark and bright sirens as independent (as in current literature), we obtain $H_0=\Hbdind$ and $\Omega_m=\Obdind$, these values are still consistent with the simulated ones. However, we still notice a shift of the posterior to low values of $H_0$ and high values of $\Omega_{\rm m}$ inherited from the wrong reconstruction of the mass model for dark sirens.
 
\begin{figure}
    \centering
    \includegraphics[width=1.0\linewidth]{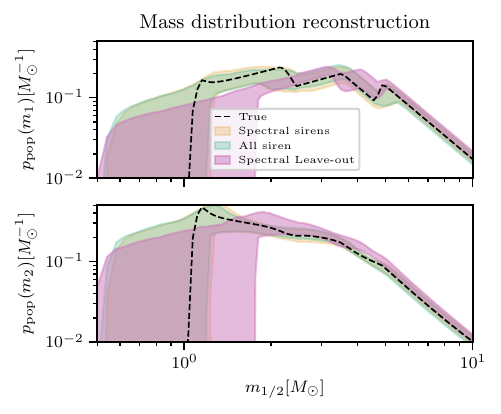}
    \caption{Reconstructed mass spectrum (90\% C.I.) for the primary mass (top panel) and secondary mass (bottom panel), three cosmological and population analyses. The analyses that consider all the sirens as dark, and bright and dark sirens with the correct framework, reconstruct the simulated mass spectrum. The analysis that uses only the dark sirens underestimates the low-mass population.}
    \label{fig:mass_reconstruction}
\end{figure}
 
In Fig.~\ref{fig:mass_reconstruction} we show the reconstructed \ac{CBC} mass spectrum in the \ac{NS} mass range. The analysis that removes all the bright sirens underestimates the \ac{NS} merger rate. The motivation for this result is that we are systematically removing low-mass \ac{GW} events as they are bright sirens, and if we do not take into account this additional selection bias, we will reconstruct a mass spectrum biased to higher masses. The level of bias on the mass spectrum and the cosmological parameters might change according to the relative fraction and types of bright and dark sirens. In this sense, we advise all future cosmological analyses that assume dark and bright sirens as independent to control the impact of this assumption.
 
\section{Conclusion}
 
In this Letter, we have reconsidered the role of bright and dark sirens for GW cosmology. By simulating a realistic population of bright and dark sirens observable in O5, we find that the role of bright sirens will be to break the degeneracy between $\Omega_m$ and $H_0$ that is observable for the dark sirens. We argue that, if additional priors on $\Omega_m$ are set from complementary observations, then bright sirens are not required to approach the level of precision required to solve the $H_0$ tension between local and early Universe \cite{Planck:2018vyg,Riess:2021jrx}.
 
We also show that, if bright and dark sirens are considered as independent samples, then it is likely that the \ac{CBC} mass spectrum will be wrongly reconstructed in its low-mass end, with the severity of the bias that can be dependent on the priors used for the population. Instead, we found that the marginal posteriors on the cosmological parameters are not strongly impacted. As such, current systematics introduced in current GW cosmology from this choice can be sub-dominant with other systematics such the the modeling of the mass spectrum \cite{Pierra:2024fbl}. However, this final result might depend on the underlying population properties of compact objects and a proper systematic study of the assumption that bright and dark siren are independent should be considered.
 
Future works can focus on exploring how \ac{EM}-related population parameters can impact the cosmological inference as well as including the information from the light-curve modelling of the \ac{GRB} and \ac{KN} emissions.

The results obtained in this Letter,
%introduced the first EM-driven unified approach for \ac{GW} cosmology with bright and dark sirens. The newly presented method frames dark and bright sirens as part of a single population, and it allows for properly including \ac{GW} events for which there are uncertainties on the actual nature of one of the two bodies. 
are obtained with a novel framework that allows us to include additional information on the intrinsic and extrinsic parameters of the binaries from the \ac{EM} emission model.
The approach can also be used for multi-messenger astronomy to infer population parameters related to the \ac{EM} emission such as the \ac{GRB} opening angle and others. 
%In this work, we considered these parameters to be perfectly known, as we were mostly interested in considering the impact of the likelihood model on the population and cosmological parameters related to the \ac{CBC}s.

%TC:ignore
\section{Acknowledgments}
We thank Maxime Bloch for comments during the internal LIGO-Virgo-KAGRA review. We are grateful to Hsin-Yu Chen for feedback and discussion about \ac{EM} selection effects.
This work is supported by ERC grant GravitySirens  101163912. Funded by the European Union. Views and opinions expressed are however those of the author(s) only and do not necessarily reflect those of the European Union or the European Research Council Executive Agency. Neither the European Union nor the granting authority can be held responsible for them. This material is based upon work supported by NSF's LIGO Laboratory which is a major facility fully funded by the National Science Foundation.

\bibliographystyle{ieeetr}
\bibliography{ref}

\section*{End Matter}

\subsection{Counterpart modeling} \label{Sec:EM_model}
\subsubsection{EM emission and ejecta properties}

To associate \ac{EM} counterparts with the simulated \ac{CBC} population, we adopt the semi-analytical framework of Ref.~\cite{Colombo:2025sdm}. For each binary, we calculate the dynamical ejecta mass $m_{\rm dyn}$ \cite{Kruger2020}, accretion disk mass $m_{\rm disk}$ \cite{barbieri2021,foucart2018}, and characteristic ejecta velocity $v_{\rm dyn}$ \cite{Radice2018} using fitting formulae calibrated to numerical-relativity simulations. Separate prescriptions are adopted for \ac{BNS} and \ac{NSBH} mergers. For \acp{BNS}, the fits depend on the component masses and tidal deformabilities $(m_1,m_2,\Lambda_1,\Lambda_2)$, with the latter determined by the \ac{NS} \ac{EoS}. We adopt the SFHo \ac{EoS}, corresponding to a maximum nonrotating \ac{NS} mass $M_{\rm NS,max}=2.06\,M_\odot$. For \acp{NSBH}, the fits additionally depend on the \ac{BH} spin, while its tidal deformability is set to zero. The ejecta properties are therefore fully specified by the intrinsic binary parameters ($m_1$, $m_2$, $\Lambda_1$, $\Lambda_2$, $\chi_{\rm BH}$), but with distinct fitting formulae reflecting their different merger dynamics. 

Unlike the \ac{KN} emission, for which the only requirement is the presence of unbound or disk material (i.e. $m_{\rm dyn}>0$ or $m_{\rm disk}>0$), the production of prompt GRB emission may require additional physical conditions. In addition to requiring a non-negligible accretion disk, we impose the formation of a \ac{HMNS}. This assumption reflects the challenges associated with launching a relativistic jet from a proto-neutron-star central engine \citep{ciolfi2020,ciolfi2020b}. We model this condition by requiring the remnant mass to satisfy $M_{\rm rem} > 1.2 \,M_{\rm NS,max}$ (see \cite{Colombo:2022zzp,Salafia:2022xjd}). Furthermore, we require the jet energy to exceed a critical threshold, following \cite{duffell2018}, ensuring that the jet can successfully break out of the surrounding ejecta and produce observable prompt emission.

The ejecta properties provide the physical input to compute the observable properties of the EM counterparts, following the method described in \cite{Colombo:2025sdm} and based on \cite{Colombo:2022zzp, Colombo:2023une}. For each binary, we compute the peak KN apparent AB magnitude in the $g$ band using an anisotropic three-component ejecta configuration, following \cite{perego2017,Breschi:2021tbm}. This approach accounts for the angular dependence of the emission and for the contribution of distinct ejecta components with different physical properties. 

The GRB prompt emission is computed using a semi-phenomenological model based on \cite{barbieri2019,Salafia:2019off}, which maps the jet energetics into an observable bolometric fluence. For BNS systems seen within a viewing angle $\theta_\mathrm{v}\leq 60^\circ$, we also include a cocoon shock breakout component, computed following the characteristics of GRB 170817A \citep{Abbott2017_GRB170817A}.  This component is not considered for \ac{NSBH} systems because of the lack of observations for these binaries. In both populations, we assumed a jet angular structure motivated by that of GRB 170817A \citep{ghirlanda2019}, characterized by a uniform core with a half-opening angle of $\theta_j = 3.4^\circ$, surrounded by wings with power-law decreasing energy
density and Lorentz factor. Further details on the semi-analytical model can be found in \cite{Colombo:2022zzp,Colombo:2023une,Colombo:2025sdm}.

\subsubsection{EM detection}
To evaluate the detectability of EM counterparts, we adopt simple threshold-based criteria on observable quantities, following the approach of \cite{Colombo:2025sdm}, without explicitly modeling a specific observational facility.

For KNe, we impose a limiting magnitude in the $g$ band of $m_{g,\mathrm{lim}} = 26$, corresponding to the depth achievable by the Vera C. Rubin Observatory \citep{Andreoni:2021epw} with an exposure time of approximately 180 s. A KN is considered detectable if its peak apparent magnitude is brighter than this threshold. While this approach neglects observational complexities such as sky localization and survey strategy, it provides a reasonable approximation for population studies. More realistic search strategies may alter the absolute detection rates by a factor of order $\sim 2$ \citep{Loffredo:2024gmx}, which we defer to future work.

For the GRB prompt emission, we adopt a fiducial fluence threshold of $F^{\rm GBM}_\mathrm{lim} = 3 \times 10^{-7}\,{\rm erg\,cm^{-2}}$, calibrated in \cite{Colombo:2025sdm} through a comparison between the predicted fluence distribution and that observed by \textit{Fermi}/GBM. We also consider a second threshold corresponding to a sensitivity improved by a factor of 10 $F^{\rm GBM\times10}_\mathrm{lim} = 3 \times 10^{-8}\,{\rm erg\,cm^{-2}}$, to explore the impact of improved sensitivity without referring to a specific future instrument.

\begin{figure}
\centering
\includegraphics[width=1.\linewidth]{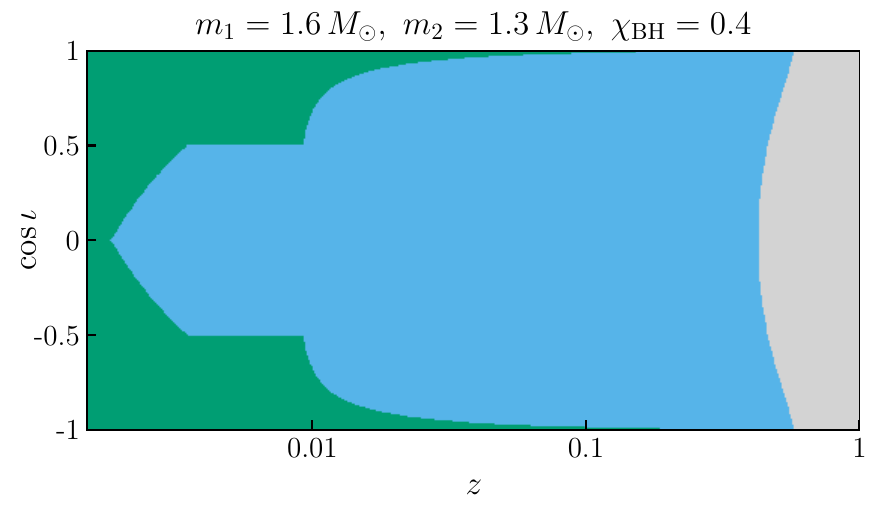}
\caption{EM detectability in the $(z, \cos\iota)$ plane, assuming  $\chi_{\rm BH}=0.4$, $m_1=1.6 M_\odot$, $m_2=1.3 M_\odot$ and the SFHo \ac{EoS}. Colors follow Figure \ref{fig:pdets}.} \label{fig:pdets_cosiota_z}
\end{figure}
 
Figure~\ref{fig:pdets_cosiota_z} shows the dependence of the selection function on the extrinsic binary parameters. Because of its collimated emission, the \ac{GRB} is detectable only for nearly face-on or face-off binaries, whereas the \ac{KN} is more isotropic but limited to lower redshifts by its apparent magnitude. A source capable of producing a counterpart can therefore enter the likelihood as either bright or dark depending on its orientation, distance, and the adopted detection threshold. The joint-detection region at $z\lesssim0.01$ is shaped by the shock-breakout component calibrated to GRB~170817A.

\begin{figure}
    \centering
    \includegraphics[width=\linewidth]{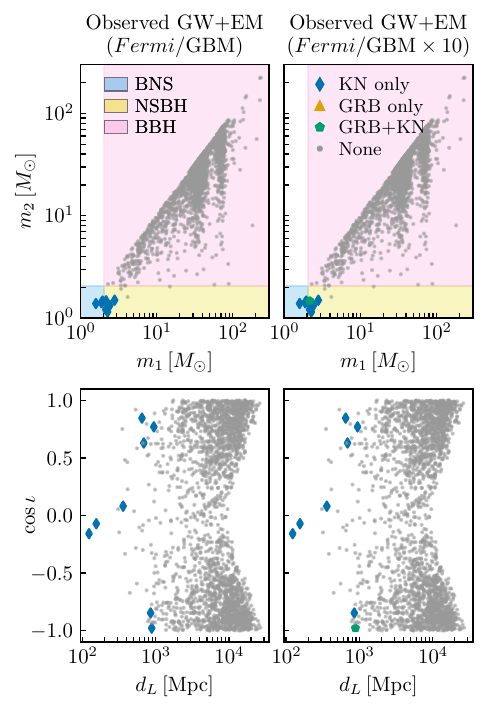}
    \caption{\textit{Top panels:} Detected bright and dark standard sirens over one year of observation in the $(m_1, m_2)$ plane. A full LVKI network operating at O5 design sensitivity is assumed, together with the SFHo EoS for NSs. The left panel adopts the nominal sensitivity of \textit{Fermi}/GBM, while in the right panel the GRB sensitivity is increased by a factor of 10. For KNe, a limiting magnitude of 26 is assumed. The shaded regions indicate the nature of the binary: pink for BBH, yellow for NSBH, and light blue for BNS systems. The markers identify the detectable EM counterparts according to our model: blue diamonds denote KN-only detections, orange triangles GRB-only detections, green pentagons joint GRB+KN detections, and grey circles events with no detectable counterpart (dark sirens). \textit{Bottom panels:} Same as the top panels, but shown in the $(\cos\iota, d_L)$ plane.}
    \label{fig:simulation_MD1}
\end{figure}

Figure~\ref{fig:simulation_MD1} visualizes the detected population used in the inference described in the main text. The bright events occupy the low-mass region containing at least one \ac{NS}, illustrating why removing them from the dark siren sample without modeling the corresponding selection biases the reconstructed compact object mass spectrum.

\clearpage
\onecolumngrid
\begin{center}
\textbf{\large Supplemental Material}
\end{center}
\onecolumngrid

\setcounter{section}{0}
\setcounter{equation}{0}
\setcounter{figure}{0}
\setcounter{table}{0}

\renewcommand{\thesection}{\Roman{section}}
\renewcommand{\theequation}{S\arabic{equation}}
\renewcommand{\thefigure}{S\arabic{figure}}
\renewcommand{\thetable}{S\arabic{table}}

\renewcommand{\theHsection}{supp.\Roman{section}}
\renewcommand{\theHequation}{supp.\arabic{equation}}
\renewcommand{\theHfigure}{supp.\arabic{figure}}
\renewcommand{\theHtable}{supp.\arabic{table}}

\input{supplement_main_text}

%TC:endignore
\end{document}

%% file: acronyms.tex
\acrodef{BBH}[BBH]{binary black hole}
\acrodef{BH}[BH]{black hole}
\acrodef{BNS}[BNS]{binary neutron star}
\acrodef{CBC}[CBC]{compact binary coalescence}
\acrodef{DAG}[DAG]{direct acyclic graph}
\acrodef{EM}[EM]{electromagnetic}
\acrodef{EoS}[EoS]{equation of state}
\acrodef{GRB}[GRB]{gamma-ray burst}
\acrodef{GW}[GW]{gravitational wave}
\acrodef{GWTC-4.0}[GWTC-4.0]{Gravitational Wave Transient Catalog}
\acrodef{KN}[KN]{kilonova}
\acrodef{HBI}[HBI]{hierarchical Bayesian inference}
\acrodef{LVK}[LVK]{LIGO-Virgo-KAGRA}
\acrodef{NS}[NS]{neutron star}
\acrodef{NSBH}[NSBH]{neutron star--black hole}
\acrodef{PE}[PE]{parameter estimation}
\acrodef{SNR}[SNR]{signal-to-noise ratio}
\acrodef{sGRB}[sGRB]{short gamma-ray burst}
\acrodef{GWTC-5.0}[GWTC-5.0]{Gravitational Wave Transient Catalog}
\acrodef{HMNS}[HMNS]{hypermassive neutron star}

%% file: supplement_main_text.tex
\section{Likelihood derivation}

We assume to have observed $N_{\rm obs}$ \ac{GW} detections, generated by a $N_{\rm astro}$ population of compact objects. We detected these source as $N_{\rm b}$ bright sirens and $N_{\rm d}$ dark sirens and we let $N_{\rm obs}=N_{\rm b}+N_{\rm d}$ to enter into the analysis. Our aim is to infer a set of population and cosmological parameters from these detections.

For each \emph{Bright Siren} indicated with index $i$, we \emph{have observed} a set of \ac{GW} data, galaxy host data and \ac{EM} data denoted with $x^i_\alpha, x^i_z, x^i_\gamma$. For these sources, we can definitely say that there was an \ac{EM} counterpart, and we will include this knowledge using a boolean random variable ${\hat{\rm EM}}$, where the hat indicates that this is set to its true state.
Moreover, for these bright sirens we also know that we have been able to \emph{detected} the GW, the host galaxy and the EM counterpart, we will include this knowledge again using a set of boolean random variables set to their true state $\mathcal{\hat{D}}^i_\alpha, \mathcal{\hat{D}}^i_z, \mathcal{\hat{D}}^i_\gamma$.
For the \emph{Dark sirens} instead, we have observed a set of GW data $\{x_\alpha\}_j$. However, we can not state if an \ac{EM} counterpart could be present or not. Further, we know that we detected the \ac{GW}, but we did not detected the host galaxy or the \ac{EM} counterpart. We will indicate this type of information on the detection and non-detection using a set of boolean variables $\mathcal{\hat{D}}^j_\alpha, \mathcal{\bar{D}}^j_z,\mathcal{\bar{D}}^j_\gamma$, where the bar indicates that the variables are set to their logical false state.

The overall likelihood can be written as 
\begin{equation}
    \mathcal{L}(N_{\rm obs},\{x_\alpha,x_z,x_\gamma,\,\hat{EM},\mathcal{\hat{D}}_\alpha,\mathcal{\hat{D}}_z, \mathcal{\hat{D}}_\gamma \}_i,\{x_\alpha,\mathcal{\hat{D}}_\alpha,\mathcal{\bar{D}}_z, \mathcal{\bar{D}}_\gamma \}_j|\Lambda,N_{\rm astro}),
\end{equation}
where $\Lambda$ indicates a set of population and cosmological parameters and the curl parenthesis indicates the set of bright and dark sirens.
% \begin{equation}
%     \mathcal{L}(N_{\rm obs},\{x_\alpha,x_z,x_\gamma\,\hat{EM},\mathcal{\hat{D}}_\alpha,\mathcal{\hat{D}}_z, \mathcal{\hat{D}}_\gamma \}_i,\{x_\alpha,\mathcal{\hat{D}}_\alpha,\mathcal{\bar{D}}_z, \mathcal{\bar{D}}_\gamma \}_j|A,B,\Gamma,H,N_{\rm astro}).
% \end{equation}
As all the \ac{GW} events are entering into the analysis, namely we are conditioning the analysis on the \ac{GW} detection, the likelihood above can be described as an inhomogeneous Poisson term \cite{Essick:2023upv,Mancarella:2024qle},
\begin{eqnarray}
    &&\mathcal{L}(N_{\rm obs},\{x_\alpha,x_z,x_\gamma,\,\hat{EM},\mathcal{\hat{D}}_\alpha,\mathcal{\hat{D}}_z, \mathcal{\hat{D}}_\gamma \}_i,\{x_\alpha,\mathcal{\hat{D}}_\alpha,\mathcal{\bar{D}}_z, \mathcal{\bar{D}}_\gamma \}_j|\Lambda,N_{\rm astro}) = \nonumber \\&& N_{\rm astro}^{N_{\rm obs}} e^{-N_{\rm astro}f(\mathcal{\hat{D}}_\alpha|\Lambda)}
    L(\{x_\alpha,x_z,x_\gamma\,\hat{EM},\mathcal{\hat{D}}_\alpha,\mathcal{\hat{D}}_z, \mathcal{\hat{D}}_\gamma \}_i,\{x_\alpha,\mathcal{\hat{D}}_\alpha,\mathcal{\bar{D}}_z, \mathcal{\bar{D}}_\gamma \}_j|\Lambda),
\end{eqnarray}
where $f(\mathcal{\hat{D}}_\alpha|\Lambda)$ is the fraction of detectable \ac{GW} sources for a population model. We can also define the number of expected \ac{GW} detections as  
\begin{equation}
    N_{\rm exp,gw}=N_{\rm astro}f(\mathcal{\hat{D}}_\alpha|\Lambda),
\end{equation}
so that the likelihood can be rearranged on a more canonical form \cite{Mandel:2018mve,Vitale:2020aaz}
\begin{eqnarray}
    &&\mathcal{L}(N_{\rm obs},\{x_\alpha,x_z,x_\gamma,\,\hat{EM},\mathcal{\hat{D}}_\alpha,\mathcal{\hat{D}}_z, \mathcal{\hat{D}}_\gamma \}_i,\{x_\alpha,\mathcal{\hat{D}}_\alpha,\mathcal{\bar{D}}_z, \mathcal{\bar{D}}_\gamma \}_j|\Lambda,N_{\rm exp,gw}) = \nonumber \\&& \frac{N_{\rm exp,gw}^{N_{\rm obs}} e^{-N_{\rm exp,gw}}}{[f(\mathcal{\hat{D}}_\alpha|\Lambda)]^{N_{\rm obs}}}
    L(\{x_\alpha,x_z,x_\gamma,\,\hat{EM},\mathcal{\hat{D}}_\alpha,\mathcal{\hat{D}}_z, \mathcal{\hat{D}}_\gamma \}_i,\{x_\alpha,\mathcal{\hat{D}}_\alpha,\mathcal{\bar{D}}_z, \mathcal{\bar{D}}_\gamma \}_j|\Lambda). 
\end{eqnarray}
Moreover, as the set of GW detection are independent from each other, we can also factorize the likelihood in terms of bright and dark siren detections as 
\begin{eqnarray}
    &&\mathcal{L}(N_{\rm obs},\{x_\alpha,x_z,x_\gamma,\,\hat{EM},\mathcal{\hat{D}}_\alpha,\mathcal{\hat{D}}_z, \mathcal{\hat{D}}_\gamma \}_i,\{x_\alpha,\mathcal{\hat{D}}_\alpha,\mathcal{\bar{D}}_z, \mathcal{\bar{D}}_\gamma \}_j|\Lambda,N_{\rm exp,gw}) = \nonumber\\&& \frac{N_{\rm exp,gw}^{N_{\rm obs}} e^{-N_{\rm exp,gw}}}{[f(\mathcal{\hat{D}}_\alpha|\Lambda)]^{N_{\rm obs}}} \prod_{i}^{N_{\rm b}} L(x^i_\alpha,x^i_z,x^i_\gamma,\,\hat{EM}^i,\mathcal{\hat{D}}^i_\alpha,\mathcal{\hat{D}}^i_z, \mathcal{\hat{D}}^i_\gamma |\Lambda)
    \prod_{j}^{N_{\rm d}}
    L(x_\alpha^j,\mathcal{\hat{D}}_\alpha^j,\mathcal{\bar{D}}_z^j, \mathcal{\bar{D}}_\gamma^j|\Lambda). 
    \label{eq:fund1}
\end{eqnarray}
The Poisson term in Eq.~\ref{eq:fund1} is the same of the likelihood model currently used for \ac{GW} cosmology \cite{Mastrogiovanni:2022ykr, Gray:2023wgj, Mancarella:2024qle, theligoscientificcollaboration2026gwtc50constraintscosmicexpansion}, and the fraction of detectable \ac{GW} sources accounts for the selection effect which is also present in the the framework currently present in literature. These terms can be evaluated using Monte Carlo integration techniques and a set of detectable \ac{GW} signals to reweight \cite{Mastrogiovanni:2023emh}. 
However, Eq.~\ref{eq:fund1} differs from the factorization of the single-event marginal likelihood terms related to the bright and dark sirens. We will see in then next sections how to factorize these terms.

\subsection{Overview of the single-event marginal likelihood}

\begin{figure}
    \centering
    \includegraphics[width=1.0\linewidth]{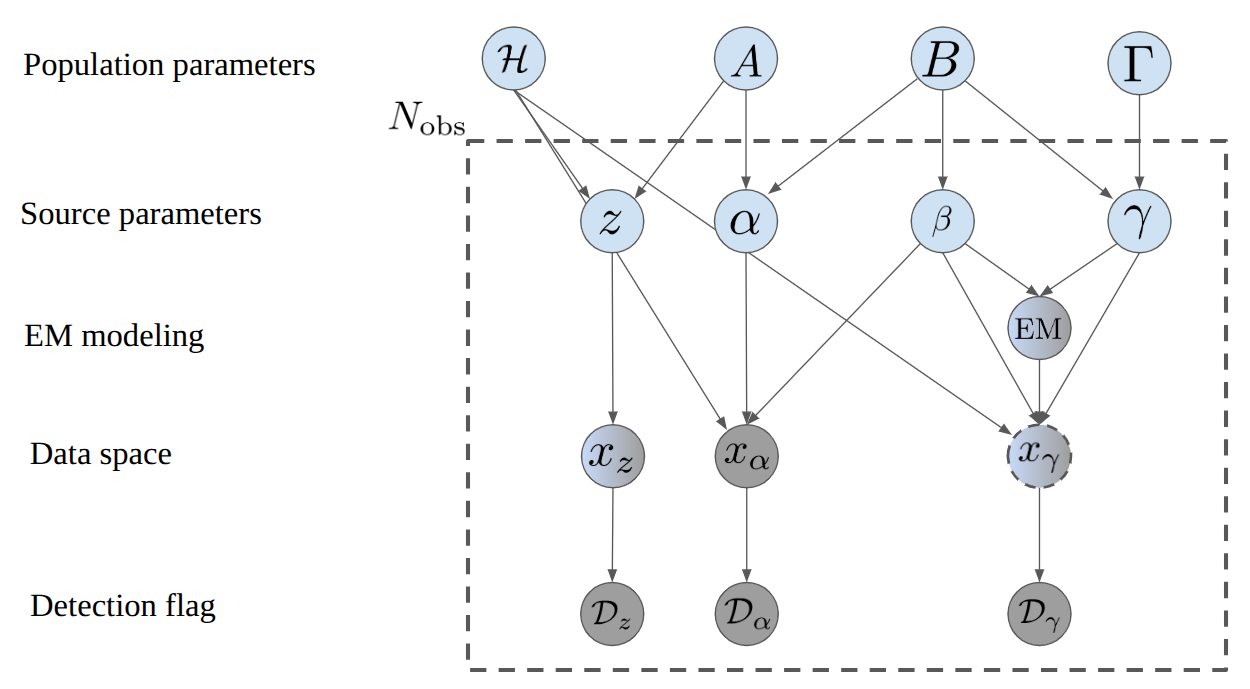}
    \caption{\ac{DAG} displaying the conditional dependencies required to model the single-event marginal likelihood. Latent variables are indicated with blue nodes while observed variables with gray nodes. Shaded nodes indicate that particular variable is observed in the bright siren case and it is latent in the dark siren case. The node indicating $x_\gamma$ (\ac{EM} data) has an additional dashed edge as in the bright siren case we can decide to ignore it and marginalize upon it \cite{Mancarella:2024qle}.  All the variables inside the dashed box have to be repeated for the $N_{\rm obs}$ \ac{GW} detections. A summary for the variables notation is given in Tab.~\ref{tab:summary}.}
    \label{fig:DAG}
\end{figure}

Before delving into the details for the derivation of the bright and dark sirens single-event marginal likelihood, let us provide an overview about the statistical dependencies and notation. The full statistical dependencies are highlighted by the \ac{DAG} in Fig.~\ref{fig:DAG}. The \ac{DAG} provides a graphic guideline to let us understand how the likelihood should be factorized. Grey nodes are observed, and cyan nodes are latent variables on which we can marginalize. If a node is shaded, it means that its observation differs from bright and dark sirens.
The single-event marginal likelihood is conditioned on a set of population parameters $\Lambda$ that we would like to infer. Moreover, the single-event marginal likelihood is marginalized over a set of a single-event latent variables that we did not observe (see Fig.~\ref{fig:DAG}). With the term ``single-event'' we indicate that these variables are proper of each source and they must considered independent draws for each source given a set of population parameters $\Lambda$. In the following, we will describe the conditional dependencies of the single-event likelihood; we refer to Tab.~\ref{tab:summary} for a summary of our notation.

We split this set of parameters as $\mathcal{H}$ relating to cosmology, $A$ related \emph{only} to the population properties of \ac{GW} sources, $\Gamma$ related \emph{only} to population properties of \ac{EM} sources and $B$ related \emph{both} to population properties of \ac{GW} and \ac{EM} sources. As an example, the Hubble constant is included in $\mathcal{H}$, the parameters describing the masses of the two objects (if not relevant to \ac{EM}) in $A$, the opening angle of the \ac{GRB}s in $\Gamma$ and the \ac{CBC} merger rate in $B$. 
The set of population parameters condition a set of source or binary parameters which are the latent variables on which the single-event marginal likelihood is marginalized upon. The connection between the population-level and binary-level parameters is encoded by a \emph{population probability} $p_{\rm pop}(\theta|\Lambda)$ which is continuous function. The binary variables are the redshift, the set of \ac{GW} binary parameters $\alpha$, the set of \ac{EM} parameters $\gamma$ and a set of source parameters $\beta$ relevant to both the \ac{EM} and \ac{GW} emission.
Below the binary-level parameters, we have the \ac{EM} modelling, which is a boolean flag indicating if an \ac{EM} counterpart is emitted or not. The possibility of emitting an \ac{EM} counterpart is a delta function that depends only on the $\beta$ and $\gamma$ parameters, namely $\delta({\rm EM}|\beta,\gamma)$.  
Below, we have the data space indicating the set of data linked to the \ac{GW}, \ac{EM} and spectroscopic observations of the galaxies. These data sets are linked to the source binary parameters with some likelihoods $\mathcal{L}(x|\cdot)$ that we typically use to measure the binary parameters from data.
Finally, we have the boolean variables for detection $\mathcal{D}$, which are always observed and are connected to their data with a delta function $\mathcal{C}(\mathcal{D}|x)$. 

\begin{table}[ht]
\centering
\begin{tabular}{ccc}
\multicolumn{2}{c}{} \\
\multicolumn{2}{c}{\textsc{Population-level parameters}} \\
\hline
\textbf{Parameter} & \textbf{Description} \\
\hline
\hline
$A$ &  Relevant only to GW parameters, minimum mass of CBC \\ 
$B$ &  Relevant to both GWs and EM, e.g. redshift merger rate slope \\
$\Gamma$ &  Relevant only to EM, e.g. GRB energy spectrum \\
$\mathcal{H}$ &  Cosmological parameters, e.g. Hubble constant \\
\hline

\multicolumn{2}{c}{} \\
\multicolumn{2}{c}{\textsc{Binary-level parameters}} \\
\hline
\textbf{Parameter} & \textbf{Description} \\
\hline
\hline
$\alpha$ &  Relevant only to GW parameters, e.g. masses \\ 
$\beta$ &  Relevant to both GWs and EM, e.g. inclination angle \\
$\gamma$ &  Relevant only to EM, e.g. peak energy GRB \\
$z$ &  Cosmological redshift \\
${\rm EM}$ &  Boolean variable indicating that the binary can be either bright or not. \\
\hline

\multicolumn{2}{c}{} \\
\multicolumn{2}{c}{\textsc{Data-level parameters}} \\
\hline
\textbf{Parameter} & \textbf{Description} \\
\hline
\hline
$x_\alpha$ &  GW data \\ 
$x_z$ &  Spectroscopic data to infer the host galaxy redshift \\
$x_\gamma$ &  EM data, e.g. GRBs light curve\\
\hline

\multicolumn{2}{c}{} \\
\multicolumn{2}{c}{\textsc{Detection-level parameters}} \\
\hline
\textbf{Parameter} & \textbf{Description} \\
\hline
\hline
$\mathcal{D}_\alpha$ &  GW detection \\ 
$\mathcal{D}_z$ &  Host galaxy detection \\
$\mathcal{D}_\gamma$ &  EM detection \\ 
\hline

\multicolumn{2}{c}{} \\
\multicolumn{2}{c}{\textsc{Connections between nodes (probabilities)}} \\
\hline
\textbf{Letter} & \textbf{Description} \\
\hline
\hline
$p_{\rm pop}(\theta|\Lambda)$ &  Continuous distribution of binary parameters conditional on population parameters. \\ 
$\mathcal{\delta({\rm EM} |\beta,\gamma)}$ &  Dirac's delta probability to describe the (non) emission of the \ac{EM} counterpart. \\ 
$\mathcal{L}(x|\theta)$ &  Likelihood function connecting binary parameters to data. \\ 
$\mathcal{C}(\mathcal{D}|x)$ &  Categorical detection or non-detection probability. \\ 
\hline

\end{tabular}
\caption{\label{tab:summary} Summary of the variables entering the hierarchical likelihood and their physical interpretation.}
\end{table}

We note that the DAG proposed in this work has different conditional dependencies from the one proposed by \cite{Essick:2026nzq}. In our approach, we assume that the probability of \ac{EM} emission and detection is conditionally dependent on the intrinsic parameters of a binary, and that when there is a non-null probability of \ac{EM} emission. While the main result of \cite{Essick:2026nzq} is obtained by assuming that the follow-up/detection is conditionally independent from \ac{EM} data given the intrinsic parameters of the binary.

\subsection{Bright sirens}
We now focus on calculating the bright sirens single-event marginal likelihood in \eqref{eq:fund1}. The likelihood can be factorized using Fig.~\ref{fig:DAG}. In writing the likelihood, we will drop the $i$ index indicating the event in order to simplify the notation. The likelihood is 
\begin{eqnarray}
    &&L(x_\alpha,x_z,x_\gamma ,\hat{EM},\mathcal{\hat{D}}_\alpha,\mathcal{\hat{D}}_z, \mathcal{\hat{D}}_\gamma |\Lambda) = \int dz d\alpha d\beta d\gamma 
    \mathcal{C}(\mathcal{\hat{D}}_z|x_z) \mathcal{C}(\mathcal{\hat{D}}_\alpha|x_\alpha) \mathcal{C}(\mathcal{\hat{D}}_\gamma|x_\gamma) \nonumber \times \nonumber \\
    && \mathcal{L}(x_z|z) \mathcal{L}(x_\alpha|\alpha,z,\beta,\mathcal{H}) \mathcal{L}(x_\gamma| {\rm \hat{EM}},\gamma,z,\beta,\mathcal{H}) \delta({\rm \hat{EM}}|\beta,\gamma)  p_{\rm pop}(z|\mathcal{H},B) p_{\rm pop}(\alpha|A,B) p_{\rm pop}(\beta|B) p_{\rm pop}(\gamma|B,\Gamma) \label{eq:bright_full}
\end{eqnarray}
In the above equation, detection probability terms $\mathcal{C}(\cdot)$ are set to $1$ as we detected the \ac{EM} counterpart. The term $\delta({\rm \hat{EM}}|\beta,\gamma)$ is restricting our integrals to a volume in the binary parameters where we are sure that an EM counterpart could be emitted. Eq.~\ref{eq:bright_full} is the full likelihood that should be used when considering information from \ac{GW} and \ac{EM} data. As an example, this is the case when want to exploit the determination of the binary parameters from \ac{GW} data and the \ac{GRB} afterglow light curves as done in \cite{Hotokezaka:2018dfi}. 

However, as noted in \cite{Chen:2023dgw, Mancarella:2024qle}, for a more \ac{EM}-agnostic inference, one can also marginalize on \ac{EM} data $x_\gamma$ \footnote{This is the motivation for which in Fig.~\ref{fig:DAG} we indicated the $x_\gamma$ node with a dashed edge}. 
We can then define
\begin{equation}
    f(\hat{EM},\mathcal{\hat{D}}_\gamma|z,\beta,B,\Gamma, \mathcal{H})= \int dx_\gamma d\gamma \, \mathcal{C}(\mathcal{\hat{D}}_\gamma|x_\gamma) \mathcal{L}(x_\gamma| {\rm \hat{EM}},\gamma,z,\beta,\mathcal{H}) \delta({\rm \hat{EM}}|\beta,\gamma)  p_{\rm pop}(\gamma|B,\Gamma),
\end{equation}
as the fraction of \ac{EM}-bright and \ac{EM}-detectable bright sirens as function of $z,\beta$ (all GW parameters) given a population model. The likelihood in Eq.~\eqref{eq:bright_full} will then become
\begin{eqnarray}
    && L(x_\alpha,x_z,x_\gamma,\,\hat{EM},\mathcal{\hat{D}}_\alpha,\mathcal{\hat{D}}_z, \mathcal{\hat{D}}_\gamma |\Lambda) = \int dz d\alpha d\beta \mathcal{L}(x_z|z) \mathcal{L}(x_\alpha|\alpha,z,\beta,\mathcal{H}) \times \nonumber \\  
    &&p_{\rm pop}(z|\mathcal{H},B) p_{\rm pop}(\alpha|A,B) p_{\rm pop}(\beta|B) f(\hat{EM},\mathcal{\hat{D}}_\gamma|z,\beta,B,\Gamma, \mathcal{H})
\end{eqnarray}
Note that the above likelihood collapses to the same likelihood in \cite{Mancarella:2024qle} in the limit that we are sure that all the \ac{GW} sources considered are emitting an \ac{EM} counterpart, and to the various cases discussed in \cite{Chen:2023dgw}. It is interesting to note that, from a statistical point of view, the case in which we detected an \ac{EM} counterpart but the \ac{EM}-associated data is uninformative on all the relevant parameters ($\mathcal{L}(x_\gamma| {\rm \hat{EM}},\gamma,z,\beta,\mathcal{H})={\rm constant}$), then the overall likelihood corresponds (apart from a normalization constant) to the one defined with the fraction of detectable sources. This is a direct consequence of the fact that, while $\mathcal{L}(x_\gamma| {\rm \hat{EM}},\gamma,z,\beta,\mathcal{H})$ can only by constant in the \ac{EM}-detectable volume and must be null for binary configurations that are not \ac{EM}-detectable.

\subsection{Dark sirens}

Differently from bright sirens, for dark sirens we know that we did not detect the host galaxy and the \ac{EM} counterpart, so in this case the redshift and \ac{EM} data becomes latent variables on which we should marginalize. More importantly, also the \ac{EM} emission flag is a latent variable as we do not know if a counterpart was emitted but not detected or not emitted at all. Referring to Fig.~\ref{fig:DAG}, the likelihood can be factorized as 
\begin{eqnarray}
    &&L(x_\alpha,\mathcal{\hat{D}}_\alpha,\mathcal{\bar{D}}_z, \mathcal{\bar{D}}_\gamma|A,B,\Gamma,\mathcal{H}) = \sum_{{\rm EM }}^{\rm \hat{EM},\bar{EM} }\int dx_z dx_\gamma dz d\alpha d\beta d\gamma \, \mathcal{C}(\mathcal{\bar{D}}_z|x_z) \mathcal{C}(\mathcal{\hat{D}}_\alpha|x_\alpha) \mathcal{C}(\mathcal{\bar{D}}_\gamma|x_\gamma) \nonumber \times \\
    && \mathcal{L}(x_z|z)\mathcal{L}(x_\alpha|\alpha,z,\beta,\mathcal{H})\mathcal{L}(x_\gamma| {\rm EM},\gamma,z,\beta,\mathcal{H}) \delta({\rm EM}|\beta,\gamma) p_{\rm pop}(z|\mathcal{H},B) p_{\rm pop}(\alpha|A,B) p_{\rm pop}(\beta|B) p_{\rm pop}(\gamma|B,\Gamma) \label{eq:dark_full}.
\end{eqnarray}
We start by marginalizing over $x_z$ by assuming that the probability of \emph{not} detecting a host galaxy at any given redshift is certain when we are observing a dark siren, 
\begin{equation}
    \int dx_z \mathcal{C}(\mathcal{\bar{D}}_z|x_z)\mathcal{L}(x_z|z) = f(\mathcal{\bar{D}}_z|z)=1.
\end{equation}
We can then integrate over $x_\gamma$, and define the fraction of \emph{non-detectable} \ac{EM} counterparts given some binary and population parameters (given also the \ac{EM} emission uncertainty),
\begin{equation}
    \int dx_\gamma \mathcal{C}(\mathcal{\bar{D}}_\gamma|x_\gamma)\mathcal{L}(x_\gamma|{\rm EM},\gamma,z,\beta,\mathcal{H}) = f(\mathcal{\bar{D}}_\gamma|{\rm EM},\gamma,z,\beta,\mathcal{H}).
\end{equation}
Finally, we note that as in the bright siren case $\mathcal{C}(\hat{D}_\alpha|x_\alpha)=1$ as we detected the \ac{GW}. It follows that the single-event marginal likelihood in Eq.~\ref{eq:dark_full} can be written as
\begin{eqnarray}
    L(x_\alpha,\mathcal{\hat{D}}_\alpha,\mathcal{\bar{D}}_z, \mathcal{\bar{D}}_\gamma|A,B,\Gamma,\mathcal{H}) = && \sum_{{\rm EM }}^{\rm \hat{EM},\bar{EM} }\int dz d\alpha d\beta d\gamma \mathcal{L}(x_\alpha|\alpha,z,\beta,\mathcal{H}) f(\mathcal{\bar{D}}_\gamma|{\rm EM},\gamma,z,\beta,\mathcal{H}) \delta({\rm EM}|\beta,\gamma) \times \nonumber \\ && p_{\rm pop}(z|\mathcal{H},B) p_{\rm pop}(\alpha|A,B) p_{\rm pop}(\beta|B) p_{\rm pop}(\gamma|B,\Gamma) \label{eq:dd}.
\end{eqnarray}
Now we will proceed for the two cases, one corresponding to systems
that emit an EM counterpart, and one for systems that do not.

In the case that an EM counterpart is \emph{not} emitted ($\bar{EM}$), then $f(\mathcal{\bar{D}}_\gamma|{\rm \bar{EM}},\gamma,z,\beta,H)=1$, as we are sure that we will never detect an EM counterpart. Therefore we can integrate on $\gamma$ and obtain
\begin{eqnarray}
L(x_\alpha,\mathcal{\hat{D}}_\alpha,\mathcal{\bar{D}}_z,
\mathcal{\bar{D}}_\gamma
|A,B,\Gamma,\mathcal{H},\bar{\rm EM})
&=&
\int dz\,d\alpha\,d\beta\,
\mathcal{L}(x_\alpha|\alpha,z,\beta,\mathcal{H})
f(\bar{\rm EM}|\beta,B,\Gamma)
\nonumber\\
&&{}\times p_{\rm pop}(z|\mathcal{H},B)
p_{\rm pop}(\alpha|A,B)
p_{\rm pop}(\beta|B)
\label{eq:nemp}
\end{eqnarray}
where we have defined the fraction of non-EM-emitting sources  
\begin{equation}
    f({\rm \bar{EM}}|\beta,B,\Gamma)=\int d\gamma \delta(\bar{EM}|\gamma,\beta)p_{\rm pop}(\gamma|B,\Gamma).
\end{equation}
Note that, if we are sure that an EM counterpart is never emitted, regardless of the parameters $\beta$, and the population parameters $B, \Gamma$, then $f({\rm \bar{EM}}|\beta,B,\Gamma)=1$ and the likelihood reduces to the standard dark siren likelihood.
   
In this case on an emitted \ac{EM} counterpart ($\hat{EM}$), $\delta(\rm {\hat{EM}}|\gamma, \beta)$ will restrict the integral in $d \gamma$  over the binary parameters that we believe will emit an \ac{EM} counterpart and $f(\mathcal{\bar{D}}_\gamma|{\rm \hat{EM}},\gamma,z,\beta,\mathcal{H})$ will provide an additional weight telling us what is the fraction of non-detectable sources emitting an EM counterpart. We can define
\begin{equation}
    f(\mathcal{\bar{D}}_\gamma,{\rm \hat{EM}}|\beta,B,\Gamma,\mathcal{H},z)=\int d\gamma  f(\mathcal{\bar{D}}_\gamma|{\rm \hat{EM}},\gamma,z,\beta,\mathcal{H}) \mathcal{\delta}(\hat{EM}|\gamma,\beta)p_{\rm pop}(\gamma|B,\Gamma),
\end{equation}
as the fraction of \emph{emitting but non-EM-detectable} binaries so that the term in the sum is 
    \begin{eqnarray}
    L(x_\alpha,\mathcal{\hat{D}}_\alpha,\mathcal{\bar{D}}_z, \mathcal{\bar{D}}_\gamma|A,B,\Gamma,\mathcal{H},\hat{\rm EM}) = && \int dz d\alpha d\beta \, L(x_\alpha|\alpha,z,\beta,\mathcal{H})f(\mathcal{\bar{D}}_\gamma,{\rm \hat{EM}}|\beta,B,\Gamma,\mathcal{H},z) \times \nonumber \\ && p_{\rm pop}(z|\mathcal{H},B) p_{\rm pop}(\alpha|A,B) p_{\rm pop}(\beta|B). \label{eq:emp}
\end{eqnarray}

By combining Eq.~\eqref{eq:emp}-\eqref{eq:nemp} into Eq.~\eqref{eq:dd}, we obtain the expression for the single-event marginal likelihood,
\begin{eqnarray}
    L(x_\alpha,\mathcal{\hat{D}}_\alpha,\mathcal{\bar{D}}_z, \mathcal{\bar{D}}_\gamma|A,B,\Gamma,\mathcal{H}) = && \int dz d\alpha d\beta \, \mathcal{L}(x_\alpha|\alpha,z,\beta,\mathcal{H}) p_{\rm pop}(z|\mathcal{H},B) p_{\rm pop}(\alpha|A,B) p_{\rm pop}(\beta|B) \times \nonumber \\ &&  \left[f({\rm \bar{EM}}|\beta,B,\Gamma)+f(\mathcal{\bar{D}}_\gamma,{\rm \hat{EM}}|\beta,B,\Gamma,\mathcal{H},z)\right] \label{eq:dark_final}
\end{eqnarray}

\subsection{The bright and dark siren likelihood}
We are now ready to combine everything in Eq.~\ref{eq:fund1}. If we consider the case for which we want to marignalize on \ac{EM} data for dark sirens, then we have
\begin{eqnarray}
    &&\mathcal{L}(N_{\rm obs},\{x_\alpha,x_z,x_\gamma,\,\hat{EM},\mathcal{\hat{D}}_\alpha,\mathcal{\hat{D}}_z, \mathcal{\hat{D}}_\gamma \}_i,\{x_\alpha,\mathcal{\hat{D}}_\alpha,\mathcal{\bar{D}}_z, \mathcal{\bar{D}}_\gamma \}_j|\Lambda,N_{\rm exp,gw}) = \frac{N_{\rm exp,gw}^{N_{\rm obs}} e^{-N_{\rm exp,gw}}}{[f(\mathcal{\hat{D}}_\alpha|\Lambda)]^{N_{\rm obs}}} \times \nonumber \\ && \prod_{i}^{N_{\rm d}}  \int dz^i d\alpha^i d\beta^i \, \mathcal{L}(x^i_\alpha|\alpha^i,z^i,\beta^i,\mathcal{H}) p_{\rm pop}(z^i|\mathcal{H},B) p_{\rm pop}(\alpha^i|A,B) p_{\rm pop}(\beta^i|B) \left[f({\rm \bar{EM}^i}|\beta^i,B,\Gamma)+f(\mathcal{\bar{D}}^i_\gamma,{\rm \hat{EM}}^i|\beta^i,B,\Gamma,\mathcal{H},z^i)\right] \nonumber \\ 
    && \prod_{j}^{N_{\rm b}}
    \int dz^j d\alpha^j d\beta^j \mathcal{L}(x^j_z|z^j) \mathcal{L}(x^j_\alpha|\alpha^j,z^j,\beta^j,\mathcal{H}) p_{\rm pop}(z^j|\mathcal{H},B) p_{\rm pop}(\alpha^j|A,B) p_{\rm pop}(\beta^j|B) f(\hat{EM}^j,\mathcal{\hat{D}}^j_\gamma|z^j,\beta^j,B,\Gamma, \mathcal{H}). 
    \label{eq:final1}
\end{eqnarray}
while instead, if we consider the case for which we want to exploit the full set of \ac{EM} data, then we have
\begin{eqnarray}
    &&\mathcal{L}(N_{\rm obs},\{x_\alpha,x_z,x_\gamma,\,\hat{EM},\mathcal{\hat{D}}_\alpha,\mathcal{\hat{D}}_z, \mathcal{\hat{D}}_\gamma \}_i,\{x_\alpha,\mathcal{\hat{D}}_\alpha,\mathcal{\bar{D}}_z, \mathcal{\bar{D}}_\gamma \}_j|\Lambda,N_{\rm exp,gw}) = \frac{N_{\rm exp,gw}^{N_{\rm obs}} e^{-N_{\rm exp,gw}}}{[f(\mathcal{\hat{D}}_\alpha|\Lambda)]^{N_{\rm obs}}} \times \nonumber \\ && \prod_{i}^{N_{\rm d}}  \int dz^i d\alpha^i d\beta^i \, \mathcal{L}(x^i_\alpha|\alpha^i,z^i,\beta^i,\mathcal{H}) p_{\rm pop}(z^i|\mathcal{H},B) p_{\rm pop}(\alpha^i|A,B) p_{\rm pop}(\beta^i|B) \left[f({\rm \bar{EM}^i}|\beta^i,B,\Gamma)+f(\mathcal{\bar{D}}^i_\gamma,{\rm \hat{EM}}^i|\beta^i,B,\Gamma,\mathcal{H},z^i)\right] \nonumber \\ 
    && \prod_{j}^{N_{\rm b}}
    \int dz^j d\alpha^j d\beta^j \mathcal{L}(x^j_z|z^j) \mathcal{L}(x^j_\alpha|\alpha^j,z^j,\beta^j,\mathcal{H}) \mathcal{L}(x^j_\beta|\beta^j, \gamma^j, \hat{\rm EM}^j,z^j,\mathcal{H}) \times 
     \nonumber \\ && p_{\rm pop}(z^j|\mathcal{H},B) p_{\rm pop}(\alpha^j|A,B) p_{\rm pop}(\beta^j|B) p_{\rm pop}(\gamma^j|\Gamma) f(\hat{EM}^j,\mathcal{\hat{D}}^j_\gamma|z^j,\beta^j,B,\Gamma, \mathcal{H}). 
    \label{eq:final2}
\end{eqnarray}

Let us discuss a few interesting limits of Eq.~\ref{eq:final1}. In the case that, we are observing all the \ac{GW}s as dark sirens, and we are sure that these objects do not emit any \ac{EM} counterpart, the hierarchical likelihood reduces to the standard likelihood used for cosmology with dark sirens \cite{Mastrogiovanni:2023zbw, Gray:2023wgj, Mancarella:2021ecn}. Instead, in the case that we are observing all the \ac{GW} sources as bright sirens, and we are sure that all of them emit an \ac{EM} counterpart, then the likelihood reduces to the case presented in \cite{Mancarella:2024qle}. Another interesting limit for the bright-siren-only case is the one for which we are sure that we are able to detect all the \ac{EM} counterparts; in that case, the likelihood reduces to the one used for bright siren analysis of GW170817 \cite{LIGOScientific:2018gmd}.

\subsection{Getting additional information for dark sirens}

\begin{figure}
    \centering
    \includegraphics[width=0.5\linewidth]{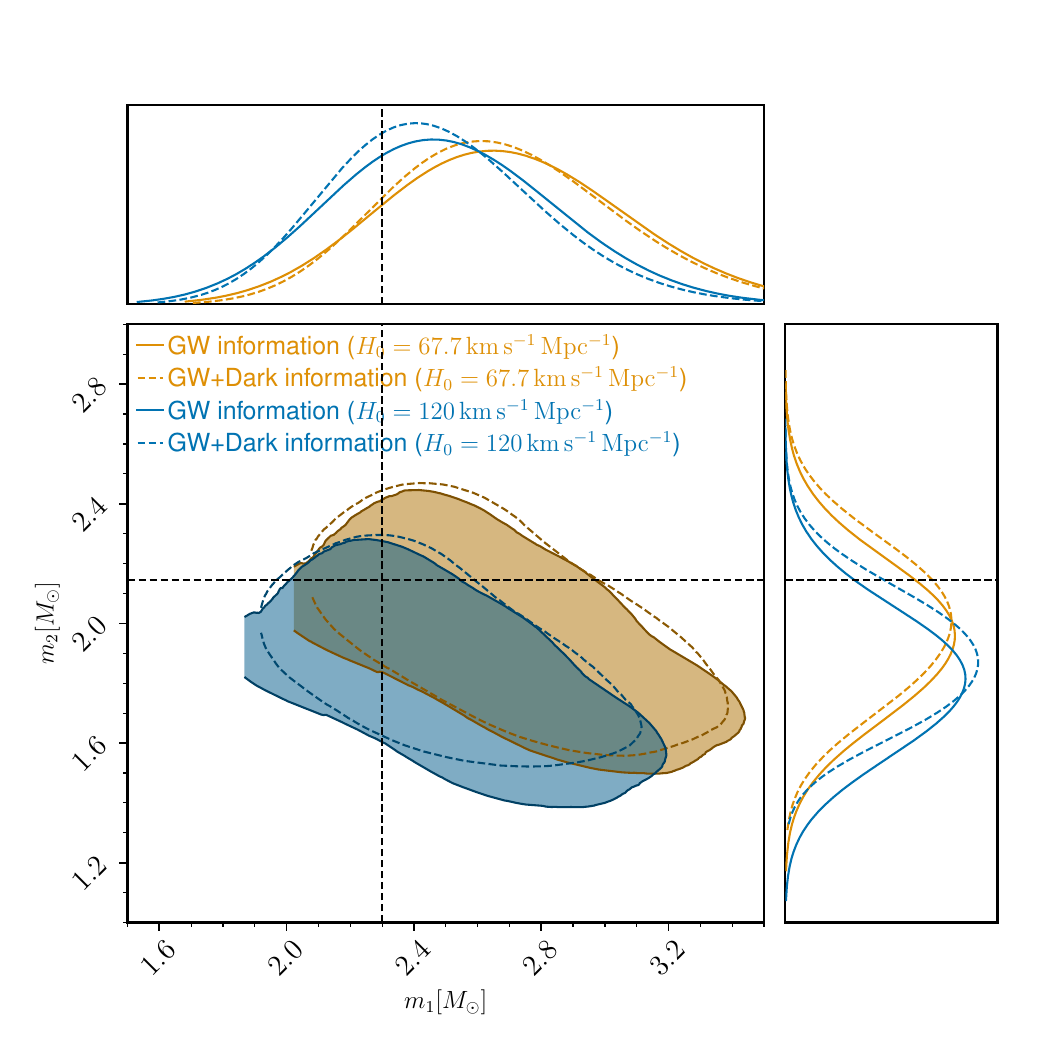}
    \caption{Posterior distributions of the source-frame component masses for a representative dark siren, under two different assumptions on the Hubble constant. The blue and orange contours correspond to $H_0 = 67.7$ and $H_0 = 120\ \mathrm{km\,s^{-1}\,Mpc^{-1}}$, respectively. Solid and dashed lines denote the inference obtained using GW information only and GW information combined with the dark siren selection effect, respectively.}
    \label{fig:dark_event_reduced}
\end{figure}

Another finding of this study is that, with a careful \ac{EM} modelling, additional cosmological information can also be included for dark sirens. Fig.~\ref{fig:dark_event_reduced} shows the posterior on the  source masses of a dark siren under the hypothesis of two different values of $H_0$. We can see that when the \ac{EM} modelling is included, certain areas of the binary mass space are excluded since the source would otherwise become \ac{EM}-bright. The improvement in precision on the mass can include additional information for the spectral siren part of the likelihood. A similar result (also present in our framework) was found in \cite{Mancarella:2024qle} for the \ac{EM} modelling of bright sirens that can help solve the degeneracy between the binary inclination angle and luminosity distance.

\section{Mock data generation}
Two datasets are needed for \ac{HBI}: a set of detectable \textit{injections} with which to estimate the selection biases, and a set of detected \textit{events} from the simulated population with which to evaluate the event-level marginal likelihood \cite{Mastrogiovanni:2023zbw}. In the following, we will refer to the first data set with the term ``\textit{injections}'', and we will refer to the latter with the term ``\textit{events}''.
Both the injections and events datasets need to be generated with the same forward modelling that identifies latent (true, physical and related to the source) variables and data (observed) variables. Typically, for toy forward models, latent and observed variables indicate the same physical quantity, \eg a true and measured mass, although observed variables are not necessarily required to have physical sense and in general they will differ. For instance, for \ac{GW} data, the observed variable is the interferometer data strain and the latent variables are the physical parameters of the system.  
Whether an event or injection is detected is a deterministic function of observed data \cite{Essick:2023upv}.

The only differences in the generation of the injection and events dataset are that \textit{(i)} injections must cover the entire detectable parameter space while events are simulated from a fiducial astrophysical model and \textit{(ii)} to evaluate the selection bias for injections we just need the detected injections and the true values of their latent variables, while to evaluate the marginal likelihood of events we need a set of posterior samples (often referred to as ``\textit{parameter estimation}'' (PE) samples) on the latent variables given the observed one.

In the following, we describe the forward modelling used to validate the all-siren methodology as well as how we generate the injection and event datasets.

\subsection{Mock likelihood}

The latent variables in our model are the primary and secondary detector masses of the binary $m_1,m_2$, the cosine of the orbital inclination angle with respect to the line-of-sight $\cos \iota$ and the source luminosity distance $d_{\rm L}$.
The total likelihood model for the forward model can be summarised as 
\begin{eqnarray}
    \mathcal{L}(A,B,P_+,P_\times,\rho^2_{\rm det}|m_1,m_2,\cos \iota,d_L) =&& \mathcal{L}(A|m_1,m_2,\rho^2_{\rm det}) \mathcal{L}(B|m_1,m_2,\rho^2_{\rm det}) \times \nn \\&& \mathcal{L}(P_+|d_L,\cos\iota,\rho^2_{\rm det}) \mathcal{L}(P_\times|d_L,\cos\iota,\rho^2_{\rm det}) \times \nn \\&& \mathcal{L}(\rho^2_{\rm det}|m_1,m_2,d_L, \cos\iota) ,
    \label{eq:Ltot}
\end{eqnarray}
where $A,B,P_+,P_\times,\rho^2_{\rm det}$ are observed data variables, and the single likelihood terms are given by
\begin{align}
&\mathcal{L}(\rho^2_{\rm det}| m_1,m_2,d_L, \cos\iota)=\chi^2_{\rm NC}(\rho^2_{\rm det}|m_1,m_2,d_L, \cos\iota,{\rm DOF}), \label{eq:Lsnr} \\
&\mathcal{L}(A|m_1,m_2,\rho^2_{\rm det})=\mathcal{N}(A|\log[\mathcal{M}(m_1,m_2)], 0.08[8/\rho_{\rm det}]), \label{eq:LM} \\
&\mathcal{L}(B|m_1,m_2,\rho^2_{\rm det})= \mathcal{N}(B|\mathcal{\eta}(m_1,m_2), 0.022[8/\rho_{\rm det}]), \label{eq:Leta} \\
&\mathcal{L}(P_+|d_L,\cos \iota,\rho^2_{\rm det})= \mathcal{N}(P_+|(1+\cos^2\iota)/(2d_L),(d_L \rho_{\rm det})^{-2}), \label{eq:LPp}\\
&\mathcal{L}(P_\times|d_L,\cos \iota,\rho^2_{\rm det})= \mathcal{N}(P_\times|\cos\iota/d_L,(d_L \rho_{\rm det})^{-2}) \label{eq:LPc}. 
\end{align}
Before we proceed with the description of the likelihoods, let us notice that we have deliberately used a notation for the data variables that does not recall any of the latent physical variables, although the likelihood models are often proposed to approximate the measure of \eg masses or other physical quantities. We used this notation to underline the important aspect that we should never confuse observed and physical variables, as we might introduce artificial biases in our inference with mock data.

The likelihood in Eq.~\eqref{eq:Lsnr} gives the probability of observing an event with squared \ac{SNR} $\rho^2_{\rm det}$ in data. The squared \ac{SNR} is used as a metric for detection. If $\rho^2_{\rm det}>12$, we detect the injection or the event. We model this likelihood as a non-central $\chi^2$ distribution with a number of degrees of freedom equal to twice the number of GW detectors available and a non-centrality parameter equal to the square optimal \ac{SNR}
\begin{equation}
    \rho = 8 \left[\frac{1+\cos^4\iota+6\cos^2\iota}{4}  \frac{1}{2} \right]^{1/2} \left[\frac{d^H_L(M)}{d_L} \right] .
    \label{eq:snrmod}
\end{equation}
The first term in the optimal \ac{SNR} quantifies the signal's power in the $+,\times$ polarisations and the second term the different sensitivities as a function of the total detector mass of the system $M=m_1+m_2$. A face-on binary ($\cos \iota=1$) with mass $M$ located at its detection horizon $d_L=d_L^H(M)$ would have \ac{SNR} 8. 
As a function for $d^H_L(M)$, we use the \ac{GW} detection horizon calculated for a Hanford, Livingston, Virgo, KAGRA and LIGO India detector network at design sensitivities to that of O5 \cite{KAGRA:2013rdx}. To generate this figure of merit, we use \textsc{gwfish} \cite{Dupletsa:2022scg} that calculates the detection horizon as the distance at which an optimally oriented binary with a certain detector mass would be detectable.

The likelihoods in Eqs.~\eqref{eq:LM}-\eqref{eq:Leta} quantify the measurement from data of the detector chirp mass and symmetric mass ratio; these likelihoods were used in previous analyses such as \cite{Fishbach:2019ckx, Mastrogiovanni:2022ykr, Farah:2023vsc, Ferraiuolo:2025evh}.
Finally, the likelihoods in Eq.~\eqref{eq:LPp}-\eqref{eq:LPc} quantify the measurement of the $+$ and $\times$ polarisations. These two likelihood terms naturally take into account the degeneracy between the determination of the luminosity distance and the inclination angle, and they correspond to a first-order Taylor expansion of the GW full likelihood around the signal's true luminosity distance and inclination angle \cite{Cutler:1994ys, Chassande-Mottin:2019nnz}. The use of these likelihoods to include the $d_L-\cos\iota$ degeneracy differs from the choice made in previous studies \cite{Fishbach:2019ckx, Mastrogiovanni:2022ykr, Farah:2023vsc, Ferraiuolo:2025evh} that used a generic ``projection factor'' that could not be directly linked to the inclination of the orbital plane (which is relevant for the prompt \ac{GRB} geometry).

\subsection{Generation of the injection set}\label{sec:injection_set}

The injection set is generated by drawing detector masses, luminosity distance and the cosine of the inclination angle for prior distributions wide enough to cover all the detectable parameter space. The luminosity distance is drawn from a distribution $\pi_{\rm inj}(d_L) \propto d_L^{1.6}$ up to a maximum distance of 32 Gpc, the cosine of the inclination angle from a uniform prior distribution $\in [-1,1]$, while the primary mass is drawn from a prior $\pi_{\rm inj}(m_1) \propto m_1^{-3}$ $\in [1,500] M_\odot$ and the secondary mass from a conditional prior distribution $\pi_{\rm inj}(m_2|m_1) \propto m_2$ defined in $\in [1,m_1] M_\odot$. The power law exponents are chosen in such a way to a set of detected injections ``similar'' to the simulated population models for the events (see Sec.~\ref{sec:evs}) in order to evaluate the selection biases with a large number of effective injections.
The total drawing prior (the product all the injection priors) is then stored as it is needed to calculate the selection biases \cite{Mastrogiovanni:2023zbw}. 

For each injection, we calculate the optimal \ac{SNR} using Eq.~\ref{eq:snrmod} and then we draw the observed \ac{SNR} according to the likelihood in Eq.~\eqref{eq:Lsnr}. If the detected \ac{SNR} exceeds 12, we store the injection as detected alongside its true binary parameters and drawing prior. The generation of detected injections is stopped when $10^5$ detected injections are reached.

\subsection{Generation of the events set}
\label{sec:evs}

Simulated \ac{GW} events are drawn in source frame and following a \ac{CBC} merger rate
\begin{equation}
    \frac{\de N_{\rm CBC}}{\de z \de m^s_1 \de m^s_2 \de \cos \iota}=T^{\rm eff}_{\rm obs} R_0 \frac{1}{1+z} \frac{dV_c}{dz}\psi(z)p_{\rm pop}(m^s_1,m^s_2)p_{\rm pop}(\cos \iota),
\end{equation}
where $V_c$ is the comoving volume, $p_{\rm pop}(m_1^s,m_2^s)$ is the \textsc{Fullpop-4.0} model in Appendix C of \cite{LIGOScientific:2025jau} with true parameters reported in Tab.~\ref{tab:priors_fullpop}, $R_0=200 {\rm yr^{-1} Gpc^{-3}}$ is the \ac{CBC} merger rate consistently with \cite{LIGOScientific:2025pvj} and $\psi(z)$ the function in Eq.~C19 \cite{LIGOScientific:2025jau} parametrizing the \ac{CBC} merger rate as a function of redshift with true parameters reported in Tab.~\ref{tab:priors_fullpop}. The total number of mergers according to the rate prescription is about $6.7 \times 10^5$ mergers per year.
The distribution of $\cos \iota$ is taken uniform, corresponding to an isotropic distribution of the orbital inclination angle. Finally, $T_{\rm obs}^{\rm eff}=0.1 {\rm yr}$ is an effective observing time chosen a posteriori to obtain $\sim 2000$ \ac{BBH} detections, that could be expected for a 5-detector network at design sensitivities in O5 \cite{Dupletsa:2022scg}. 
In other words, the effective time quantifies the fact that detectors are not always running and not all the binaries are optimally oriented. The source variables are then converted to detector variables using a $\Lambda$CDM cosmology with $H_0= 67.7   \hu, \Omega_m=0.308$ \cite{Planck:2018vyg}. 
The full mass spectrum that we choose for the simulation describes \ac{BBH}, \ac{NSBH} and \ac{BNS} encodes a preference for equal-mass binaries as well as an under density of compact objects in the $[M_{\rm NS,max}M_{\rm BH}]$ mass region, where $M_{\rm NS,max}$ is the maximum mass of \ac{NS} set to 2.06 $M_\odot$ according to the \textsc{SFHo} \ac{EoS} \cite{Hempel2012} and  $M_{\rm BH}$ is set to 5 $M_\odot$.  The CBC merger rate is chosen so that it increases before $z=2$ and then rapidly decreases to a null value. The fiducial mass and rate models are taken consistently with the ones measured from current data \cite{theligoscientificcollaboration2026gwtc50populationpropertiesmerging}, although some of the parameters, such as the redshift peak of the \ac{CBC} merger rate, are not measured. The full list of injected population parameters, as well as their prior ranges for the later inference, is reported in Tab.~\ref{tab:priors_fullpop}
\begin{table}[t]
\centering
\begin{tabular}{cccc}
\multicolumn{4}{c}{} \\
\multicolumn{4}{c}{Injected Population model} \\
\hline
\textbf{Parameter} & \textbf{Description} & \textbf{Value} & \textbf{Prior} \\
\hline
\hline
$\alpha_{1}$ & Spectral index of the power law before $b$ & $-1$ & $-1$\\
$\alpha_{2}$ & Spectral index of the power law  after $b$ & $3$ & $3$\\
$\beta_{1}$ & Spectral index of the pairing function before $m_{\rm break}$ & $2$ & $2$\\
$\beta_{2}$ & Spectral index of the pairing function after $m_{\rm break}$ & $1$ & $1$\\
$m_{\rm min}$&  Minimum primary and secondary mass [$M_\odot$]& $1.0$ & $\text{U}(0.4,1.4)^*$\\ 
$m_{\rm max}$& Maximum primary and secondary mass [$M_\odot$] & $300$ & $\text{U}(200,400)$\\ 
$\delta_{\rm m}^{\rm min}$&  1st smoothing parameter of the low mass [$M_\odot$]& $0.2$ & $0.2$\\ 
$\delta_{\rm m}^{\rm max}$& 2nd smoothing parameter of the low mass [$M_\odot$] & $10$ &$10$\\ 
$\mu_{\rm g}^{\rm low}$& Location of the first peak [$M_\odot$]& $35$ & $\text{U}(20,50)$\\ 
$\sigma_{\rm g}^{\rm low}$& Width of the first peak [$M_\odot$] & $5$ & $5$\\ 
$\mu_{\rm g}^{\rm high}$& Location of the second peak [$M_\odot$] & $70$ &$\text{U}(60,100)$\\ 
$\sigma_{\rm g}^{\rm high}$ & Width of the second peak [$M_\odot$] & $5$ & $5$\\
$\lambda_{\rm g}$& Fraction of sources in peaks & $0.05$ & $0.05$\\ 
$\lambda_{\rm g}^{\rm low}$& Fraction of sources in the first peak & $0.8$ & $0.8$\\ 
$m_{\rm d}^{\rm low}$ & Left side of the dip  [$M_\odot$]& $2.4$ & $\text{U}(1.5,3.5)^*$\\
$m_{\rm d}^{\rm high}$ & Right side of the dip [$M_\odot$]& $5$ & $\text{U}(5,9)$\\ 
$\delta_{\rm d}^{\rm min}$  & Smoothing of the left side of the dip [$M_\odot$]& $0.5$ & $0.5$\\ 
$\delta_{\rm d}^{\rm max}$  & Smoothing of the right side of the dip [$M_\odot$]& $0.5$ & $0.5$\\ 
$\rm A$  & Amplitude of the dip & $0.5$ & $\text{U}(0,1)$\\ 

\hline\hline
$\gamma$ & Slope of the power law before the point $z_{\rm p}$ & $2.7$ & $\text{U}(0,5)$ \\
$\kappa$ & Slope of the power law after the point $z_{\mathrm{p}}$ & $6$ & $6$ \\
$z_{\rm p}$ & Redshift turning point between the power laws & $2$  & $2$\\ 
$R_0$ & \ac{CBC} merger rate today in ${\rm Gpc^{-3} yr^{-1} }$ & $200$  & $-$\\ 
\hline\hline
$H_0$ & Hubble constant in \hu & $67.7$ & $\text{U}(20,140)$ \\
$\Omega_m$ & Dark matter energy fraction & $0.308$ & $\text{U}(0,1)$ \\
\hline
\end{tabular}
\caption{\label{tab:priors_fullpop}
 Summary of the hyperparameters and priors used for the population parameters. \text{U} (\text{LU}) stands for uniform (log-uniform) prior. Note$^*$: For the dark siren analysis leaving out the bright sirens, the priors on $m_{\rm min}$ and $m_{\rm d}^{\rm low}$ have been extended to $\text{U}(0.4,2.99)$ and $\text{U}(3.0,4.99)$ to explore more possible configurations of the \ac{NS} mass spectrum}
\end{table}

Once the set of \ac{GW} signals is drawn, we calculate their optimal \ac{SNR} using Eq.~\eqref{eq:snrmod}, then a detected SNR is drawn from the likelihood in Eq.~\eqref{eq:Lsnr}. If the detected \ac{SNR} exceeds a threshold of 12, then we consider the source as detected. 
Once a source is detected, we draw a set of observed parameters $A, B, P_+,P_\times$ from the likelihoods in Eq.~\eqref{eq:LM}-\eqref{eq:LPc}. We note that these ``observed'' parameters not physical, must be drawn from the unconstrained likelihoods in Eq.~\eqref{eq:LM}-\eqref{eq:LPc}, otherwise they would introduce another selection bias that should be accounted.
Finally, to obtain \ac{PE} to use for cosmological inference, we sample $\cos \iota, d_L, m_1,m_2$ (detector-frame quantities) from the \ac{GW} likelihood in Eq.~\eqref{eq:Ltot} using the observed points and \textsc{EMCEE} \cite{2013ascl.soft03002F}. In doing so, we set uniform priors in all the parameters besides a prior $p(d_L) \propto d_L^2$ on the luminosity distance.

\section{Tabulated EM emission and interpolation}
The direct evaluation of the EM emission for each binary, based on semi-analytical models, can be computationally expensive when applied to large synthetic populations. To make the analysis tractable, we adopt a tabulated approach in which the EM observables are pre-computed over a multidimensional grid of physical parameters and then evaluated through interpolation.

For the KN emission, we construct lookup tables for the peak apparent magnitude in the $g$ band as a function of the ejecta properties and viewing geometry. In particular, the tables are computed over a grid in dynamical ejecta mass $m_{\rm dyn}$, disk mass $m_{\rm disk}$, ejecta velocity $v_{\rm dyn}$, and viewing angle $\iota$. The dependence on redshift is treated at the interpolation stage. For each source, the observer-frame frequency is mapped to the corresponding rest-frame frequency according to $\nu_{\rm rest} = (1+z)\,\nu_{\rm obs}$. We note that the tabulated emission is defined at a fixed reference redshift $z_0$, so that the effective rest-frame frequency entering the interpolation is rescaled accordingly. The observed flux is then obtained using the luminosity distance $d_L(z)$, which is used to convert the intrinsic emission into an apparent magnitude. This procedure effectively accounts for cosmological redshifting and distance scaling within the interpolation scheme.

For the prompt GRB emission, we adopt a similar strategy. In this case, the relevant quantity controlling the emission is the jet core energy $E_{\rm c}$, which is determined by the remnant disk mass and the jet-launching efficiency. We therefore construct tables for the prompt bolometric fluence as a function of $E_{\rm c}$, viewing angle $\theta_{\rm v}$, and redshift $z$. These tables encapsulate both the angular structure of the jet and the mapping between intrinsic energetics and observed fluence.

In practice, for each binary in the population, we first compute the ejecta and remnant properties using the fitting formulae described above. These quantities are then used to query the precomputed tables and obtain the corresponding EM observables through multidimensional interpolation. This approach reduces the computational cost by several orders of magnitude compared to a full on-the-fly evaluation of the EM model, while accurately reproducing the behavior of the semi-analytical framework of \cite{Colombo:2025sdm}.

\section{Impact of the Neutron Star's Equation of State modelling}

The choice of the \ac{EoS} governs the transition mass between \acp{NS} and \acp{BH}, and the efficiency of the associated \ac{EM} emissions, with a direct impact on their detectability \cite{Margalit_2017,Jacobi_2023}. In our unified  framework, the change of the \ac{EoS}  directly affects the distinction between bright and dark sirens, with possible repercussions on the reconstruction of cosmological parameters. 

\begin{figure} 
\centering
\includegraphics[width=0.5\linewidth]{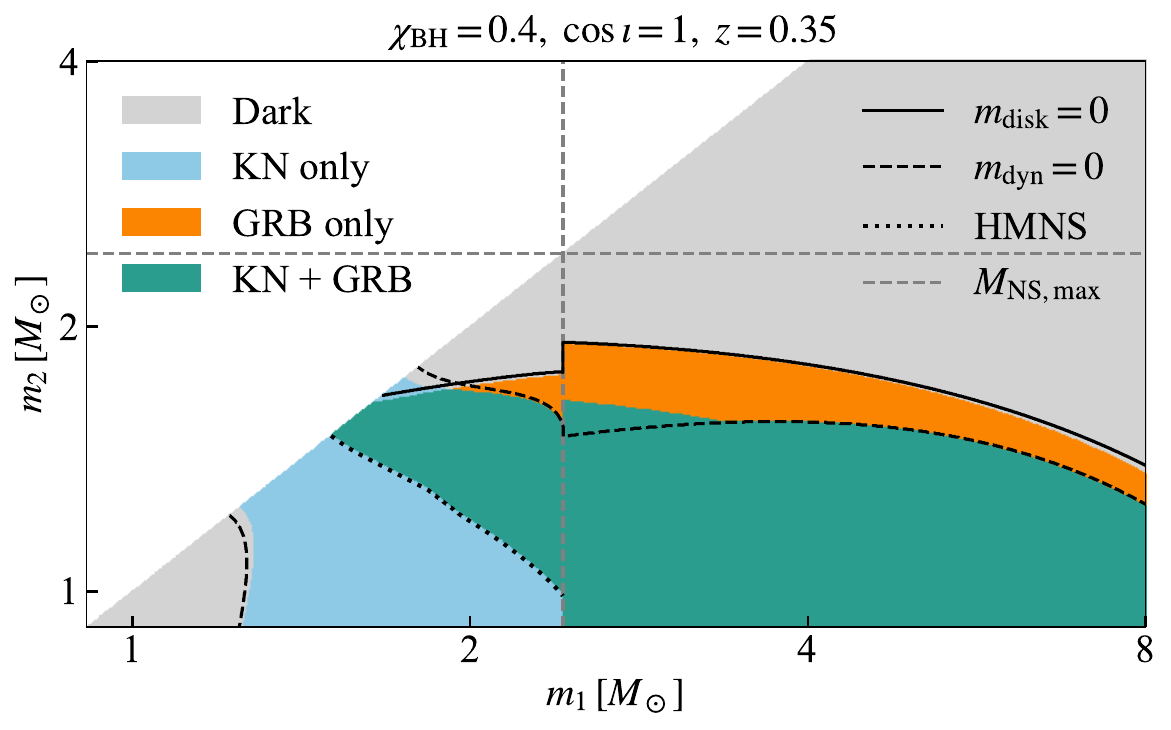} 
\caption{EM detectability in the $(m_1, m_2)$ plane, assuming  $\chi_{\rm BH}=0.4$, $\cos\iota=1$ and $z=0.35$. The colored regions indicate the type of detectable EM counterpart, grey corresponds to no detectable emission (dark), light blue to KN only, orange to GRB only, and green to joint KN+GRB detection. The grey dashed lines mark the maximum NS mass $M_{\rm NS,max}=2.42\,M_\odot$ for the DD2 EoS. The black solid, dashed, and dotted curves denote the conditions $m_{\rm disk}=0$, $m_{\rm dyn}=0$, and $m_{\rm rem}=1.2\,M_{\rm NS,max}$, respectively, the latter corresponding to the threshold for the formation of a HMNS.} 
\label{fig:pdets_dd2} 
\end{figure}

To quantify what is the effect of mismatching the \ac{EoS} model, we simulate mock data using the DD2 EoS \cite{Steiner2013} and then use the SFHo model for inference. The DD2 model yields larger stellar radii, higher tidal deformabilities, and a higher maximum mass ($M_{\rm NS,max} \approx 2.42 \, \mathrm{M}_{\odot}$) for non-rotating \acp{NS}, as shown in Figure \ref{fig:pdets_dd2}.

Under the LVKI O5 network and \textit{Fermi}/GBM$\times$10 observing scenario, the stiffer DD2 EoS yields one additional isolated GRB detection. The associated KN emission reaches a peak apparent magnitude of $m_g = 26.15$ mag, placing it immediately below our detection threshold of $m_{g,\mathrm{lim}} = 26$ mag. Notably, assuming a slightly more optimistic sensitivity for the Vera C. Rubin Observatory (e.g., $m_{g,\mathrm{lim}} = 26.53$ mag, as in \cite{Loffredo:2024gmx}), this source would be classified as a joint GW+KN+GRB detection. This event originates from a NSBH binary with primary mass of $m_1 = 4.2 \, \mathrm{M}_\odot$ and a nearly face-on orientation ($\cos\iota \approx -1$), located at a luminosity distance of $d_L \approx 1.7 \times 10^3$\,Mpc. This detection yields one additional bright siren for our cosmological analysis. In the following, we assess the propagation of these EoS effects into the all-sirens cosmological inference.

\begin{figure}[h!]
    \centering
    \subfloat[All-sirens analysis comparing DD2 and SFHo.%
    \label{fig:DD2_SFHo}]{%
        \includegraphics[width=0.5\textwidth]{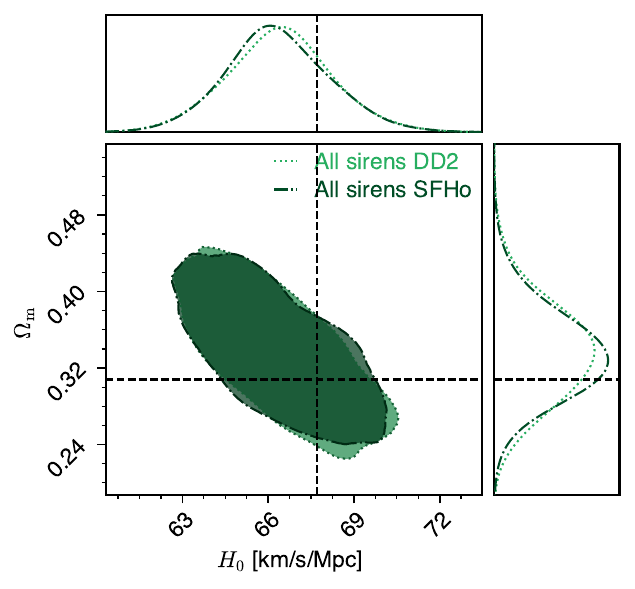}%
    }
    \subfloat[Stress-test: assumed SFHo on a DD2 mock..\label{fig:stress_test}]{%
        \includegraphics[width=0.5\textwidth]{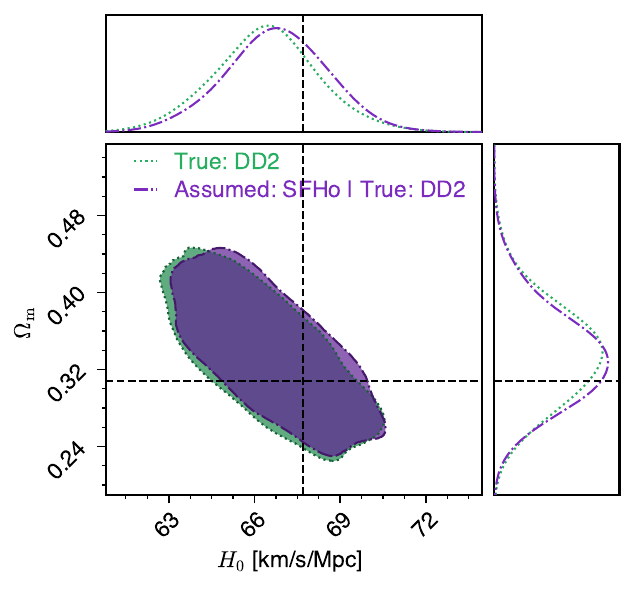}%
    }
    \caption{Posterior distributions of $\Omega_{\rm m}$ and $H_0$ obtained using the all-siren method. The contours show the joint 2D posteriors, while the marginalised 1D distributions are reported along the diagonal. The dashed black lines indicate the fiducial cosmological parameters. \textit{Panel a}: comparison of the posterior distributions obtained for the DD2 (light-green dotted line and lighter-green region) and SFHo (dark-green dash-dotted line and darker-green region) EoS. \textit{Panel b}: stress-test comparing the analysis under the true DD2 EoS (green dotted line and green region) against the parameter inference obtained by assuming an incorrect SFHo EoS on the same DD2 mock dataset (purple dash-dotted line and purple region).}
    \label{fig:EoS}
\end{figure}

First, we test the pipeline under ideal conditions, where the EoS assumed for inference matches exactly the one used to generate the mock data. As shown in Fig.~\ref{fig:DD2_SFHo}, the resulting cosmological posteriors for $H_0$ and $\Omega_{\rm m}$ obtained under DD2 assumption are fully consistent with our fiducial SFHo constraints. The single additional bright siren provides a negligible statistical advantage, confirming that the stiffness of the EoS does not alter the cosmological constraints.

Next, we perform a \textit{stress-test} to assess the risk of adopting an incorrect EoS during parameter inference.
Specifically, we evaluate a scenario where the mock data is governed by the DD2 EoS, while the cosmological pipeline assumes the SFHo EoS.
As demonstrated by the marginalized posteriors in Fig.~\ref{fig:stress_test}, the reconstructed distributions for both $H_0$ and $\Omega_{\rm m}$ exhibit no statistically significant shifts. This invariance indicates that the framework effectively decouples the nuclear physics from the cosmological parameters. Consequently, in our model an incorrect choice of the nuclear EoS does not compromise the final cosmological parameter extraction.

